\documentclass{pasj00}
\begin{document}
\SetRunningHead{Kajisawa et al.}{MOIRCS Deep Survey. IX.}
%\Received{}%{yyyy/mm/dd}
\Accepted{}%{2010/11/29}
%\Published{}%{yyyy/mm/dd}

\title{MOIRCS Deep Survey. IX.\\ Deep Near-Infrared Imaging Data and Source Catalog}

%%% begin:list of authors
% Do NOT capitalize all letters in "textsc".
\author{Masaru \textsc{Kajisawa}\altaffilmark{1,2}, Takashi \textsc{Ichikawa}\altaffilmark{2}, Ichi \textsc{Tanaka}\altaffilmark{3}, Toru \textsc{Yamada}\altaffilmark{2}, Masayuki \textsc{Akiyama}\altaffilmark{2}, Ryuji \textsc{Suzuki}\altaffilmark{3},\\ Chihiro \textsc{Tokoku}\altaffilmark{2}, Yuka Katsuno \textsc{Uchimoto}\altaffilmark{4}, Masahiro \textsc{Konishi}\altaffilmark{4}, Tomohiro \textsc{Yoshikawa}\altaffilmark{5}, Tetsuo \textsc{Nishimura}\altaffilmark{3}, Koji \textsc{Omata}\altaffilmark{3},
Masami \textsc{Ouchi}\altaffilmark{6}, Ikuru \textsc{Iwata}\altaffilmark{7}, Takashi \textsc{Hamana}\altaffilmark{8}, Masato \textsc{Onodera}\altaffilmark{9,10}}

\email{kajisawa@cosmos.ehime-u.ac.jp}

\altaffiltext{1}{Research Center for Space and Cosmic Evolution, Ehime University, Bunkyo-cho 2-5, Matsuyama 790-8577, Japan}
\altaffiltext{2}{Astronomical Institute, Tohoku University, Aramaki,
Aoba, Sendai 980--8578, Japan}
\altaffiltext{3}{Subaru Telescope, National Astronomical Observatory
of Japan, 650 North Aohoku Place, Hilo, HI 96720, USA}
\altaffiltext{4}{Institute of Astronomy, University of Tokyo, Mitaka, Tokyo
181--0015, Japan}
\altaffiltext{5}{Koyama Astronomical Observatory, Kyoto Sangyo University, Motoyama, Kamigamo, Kita-ku, Kyoto 603--8555, Japan}
\altaffiltext{6}{Observatories of the Carnegie Institution of Washington, 813 Santa Barbara Street, Pasadena, CA 91101, USA}
%\altaffiltext{6}{Institute for Cosmic Ray Research,
%University of Tokyo, 
%5-1-5 Kashiwa-no-Ha, Kashiwa City
%Chiba, 277-8582, Japan}
\altaffiltext{7}{Okayama Astrophysical Observatory, National Astronomical Observatory of Japan, Kamogata, Asakuchi, Okayama, 719--0232, Japan}
\altaffiltext{8}{National Astronomical Observatory of Japan, Mitaka, Tokyo
181--8588, Japan}
\altaffiltext{9}{Service d'Astrophysique, CEA Saclay, Orme des Merisiers, 91191 Gif-sur-Yvette Cedex, France}
\altaffiltext{10}{Institute for Astronomy, ETH Zurich, Wolfgang-Pauli-strasse 27, 8093 
Zurich, Switzerland} 

%\author{Masaru \textsc{Kajisawa} %
%  \thanks{Example: Present Address is xxxxxxx}}
%\affil{Research Center for Space and Cosmic Evolution, Ehime University, Bunkyo-cho 2-5, Matsuyama 790-8577, Japan}
%\email{kajisawa@cosmos.ehime-u.ac.jp}

%\author{Takashi \textsc{Ichikawa}, Masayuki \textsc{Akiyama}, Chihiro \textsc{Tokoku}, Toru \textsc{Yamada}}
%\affil{Astronomical Institute, Tohoku University, Aramaki,
%Aoba, Sendai 980--8578, Japan}
%
%\author{Ichi \textsc{Tanaka}, Ryuji \textsc{Suzuki}, 
%\and
%\author{C-Firstname {\sc C-Familyname}}
%\affil{C-Address of Institute}\email{ccccc@xxx.xxx.xx.xx}
%%% end:list of authors

%%% Please use the following style in case that sorting by 
%%% affilation is impossible. 
%
% \author{%
%   D-Firstname \textsc{D-Familyname}\altaffilmark{1}
%   E-Firstname \textsc{E-Familyname}\altaffilmark{1,2}
%   and
%   F-Firstname \textsc{F-Familyname}\altaffilmark{2}}
% \altaffiltext{1}{Address of Institute}
% \email{ddddd@xxx.xxx.xx.xx}
% \email{eeeee@xxx.xxx.xx.xx}
% \altaffiltext{2}{Address of Institute}

%% `\KeyWords{}' always has to be placed before `\maketitle'.
\KeyWords{catalogs --- galaxies:high-redshift --- galaxies:photometry --- infrared:galaxies --- surveys} %Do NOT move this preamble from here!

\maketitle

\begin{abstract}
%Please read ``IMPORTANT NOTICE'' carefully before preparing a manuscript. 
We present deep $J$-, $H$-, and $K_{s}$-band imaging data of the MOIRCS Deep 
Survey (MODS), which was carried out with Multi-Object Infrared Camera and 
Spectrograph (MOIRCS)  mounted on the Subaru telescope
in the GOODS-North region. 
The data reach 5$\sigma$ total limiting magnitudes for point sources 
of $J=23.9$, $H=22.8$, and $K_{s}=22.8$ 
(Vega magnitude) over 103 arcmin$^{2}$ (wide field). 
In 28 arcmin$^{2}$ of the survey area, which is  ultra deep field of 
the MODS (deep field), 
the data reach the 5$\sigma$ depths of $J=24.8$, $H=23.4$, and $K_{s}=23.8$. 
The spatial resolutions of the combined images are FWHM $\sim 0.6$ arcsec and 
$\sim 0.5$ arcsec for the wide and deep fields in all bands, respectively.
Combining the MODS data with 
the multi-wavelength public data taken with the HST, Spitzer, and other 
ground-based telescopes in the GOODS field, 
we construct a multi-wavelength photometric catalog of $K_{s}$-selected sources.
Using the catalog, we present $K_{s}$-band number counts and near-infrared color 
distribution of the detected objects, and demonstrate some selection techniques with 
the NIR colors for high redshift galaxies.   
These data and catalog are publicly available via internet.

\end{abstract}

\section{Introduction}
In recent years, several deep multi-wavelength surveys have been carried out 
to reveal galaxy formation and evolution in the high-redshift universe.
Representative examples of such surveys are the Great Observatories Origins Deep 
Survey (GOODS, \cite{gia04}), the Subaru  XMM-Newton Deep Survey (SXDS, 
K. Sekiguchi et al., in preparation), the Cosmic Evolution Survey (COSMOS, 
\cite{sco07}), and the All-wavelength Extended Groth strip International Survey 
(AEGIS, \cite{dav07}).
Deep multi-wavelength observations from radio to X-ray allow us to 
comprehensively investigate  
the properties of stars, gas, dust, and AGN of high-redshift galaxies. 
In such multi-wavelength surveys 
for high-redshift galaxies, 
near-infrared (NIR) imaging is essential for the following reasons.
First, the observed NIR luminosity 
of galaxies reflects their stellar mass, which is one of the most basic physical 
properties of galaxies,  relatively well for galaxies at $z\lesssim3$. 
NIR data also covers the Balmer/4000\AA\ break of galaxies at 
$1\lesssim z \lesssim 4$, which is important to study the stellar population of 
these galaxies and to determine their photometric redshifts. 
Compared to optical light, 
observed NIR light is less sensitive to the effect of dust extinction, and 
the spatial resolution of NIR data is comparatively good even in  
ground-based observations. 
These are helpful for the identification of sources detected 
in other wavelengths such as X-ray, mid-IR, sub-mm and radio.

The recent availability of wide-field NIR instruments with large-format detectors 
mounted on 4-8m telescopes has allowed us to carry out wider and deeper 
NIR surveys.
%Ongoing UKIRT Infrared Deep Sky Survey and Ultra-VISTA survey are 
With such new instruments, we still need a relatively long integration time 
in order to construct a representative sample of high-redshift galaxies. 
For example, a normal L$^{*}$ galaxy seen in the local universe would have 
$K_{\rm Vega} \sim 23 $ if placed at $z\sim3$ (e.g., \cite{fra03}).
Deeper data would be desirable for the investigation of 
low-mass (sub-L$^{*}$) galaxies at such high redshift. 
The search for star-forming galaxies at $z\gtrsim7$ also requires 
extremely deep NIR data.  
Existing NIR surveys with such depth are limited to several small field surveys.
They include those with the Hubble Space Telescope (HST)/NICMOS 
such as Hubble Deep Field North (HDF-N; 
\cite{dic03}; \cite{tho99}), HDF-South NICMOS field \citep{wil00},  
and Hubble Ultra Deep Field (UDF, \cite{tho05}). 
The new HST instrument, 
WFC3 IR-channel is providing deeper data than those with NICMOS 
(e.g., \cite{win10}), but the deep NIR observations  with  the HST instruments are 
practically limited 
to $\lambda < 1.8 \mu$m.  
Although 
there are also ultra-deep NIR surveys with ground-based 8m class telescopes such as 
Subaru Deep Field \citep{mai01} and Faint Infrared Extragalactic Survey HDF-S 
field (FIRES; \cite{lab03}), these surveys have small fields of several arcmin$^{2}$.  
  \begin{figure*}
  \begin{center}
    \FigureFile(140mm,140mm){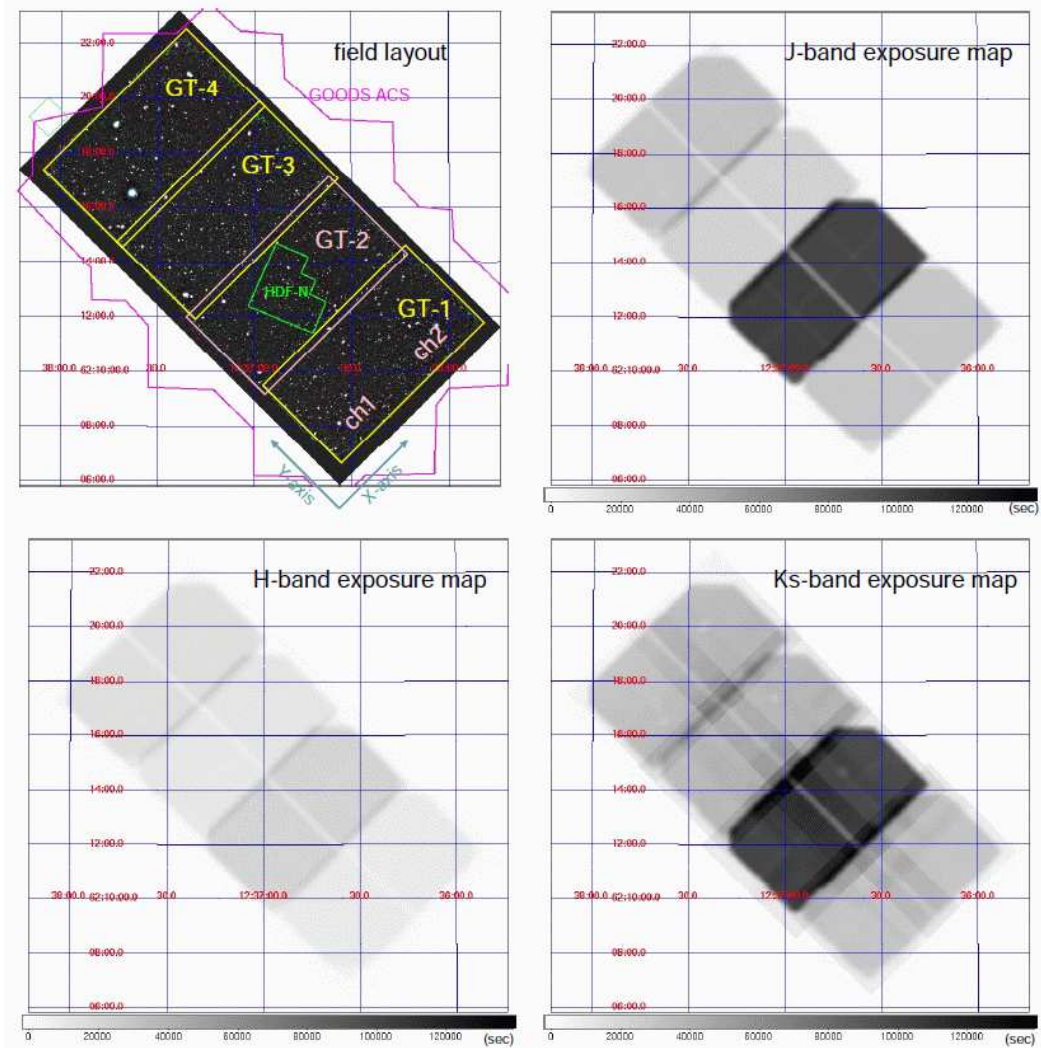}
    %%% \FigureFile(width,height){filename}
  \end{center}
\vspace{-3mm}
  \caption{Survey field layout and exposure maps for $J$, $H$, and $K_{s}$ bands.
In the top-left panel, the GOODS-N HST/ACS region and the original 
HDF-N region are shown in magenta and green lines.
Arrows represent the direction of X- and Y-axes of the mosaiced MOIRCS images.
Grayscales in the top-right, bottom-left and bottom-right panels show 
the integration time of the MOIRCS $J$, $H$, and $K_{s}$-band imaging, respectively. 
}
\label{fig:fov}
\end{figure*}

We have carried out a ultra-deep NIR imaging survey, namely, 
MOIRCS Deep Survey (MODS) with 
Multi-Object Infrared Camera and 
Spectrograph (MOIRCS)  mounted on the Subaru telescope
in the GOODS-North region. 
Total integration time of $\sim$ 124 hours were spent in the $JHK_{s}$-band imaging 
observations with MOIRCS, and the data reach $K_{s}\sim23$ (5$\sigma$, Vega) over 
$\sim$ 103 arcmin$^{2}$, and $K_{s}\sim24$ over $\sim$ 28 arcmin$^{2}$. 
In this region, the GOODS and the Chandra Deep Field North (CDF-N) surveys 
provided deep multi-wavelength imaging data  such as optical HST/ACS images 
\citep{gia04}, mid-IR images obtained with Spitzer/IRAC and MIPS 
(M. Dickinson et al., in preparation), and Chandra X-ray images \citep{ale03}.
Extremely deep radio observations with VLA,  deep sub-mm/mm 
surveys with SCUBA, AzTEC and MAMBO,  
and  extensive optical spectroscopic surveys with the Keck telescopes  
have also been carried out (e.g., \cite{mor10}; \cite{pop06}; \cite{gre08}; 
\cite{per08}; \cite{bar08}; \cite{wir04}).  
On the other hand,  NIR data which cover the GOODS-North region 
have reached only $m_{\rm AB} \sim $ 22--22.5 (\cite{cap04}; \cite{bun05}),   
although \citet{wan10} recently 
published relatively deep $K_{s}$-band data which reach $K_{s}\sim22.7$ (Vega). 
For $J$ band, there had been  no wide-field data 
which cover most of the GOODS-N region.  
 Therefore NIR data with a comparable depth have been desirable in this field.
In this context, 
with the obtained ultra-deep NIR data, we have investigated the number counts 
of Distant Red Galaxies at $z\gtrsim2$ \citep{kaj06}, the clustering properties 
of stellar mass-selected sample at $1<z<4$ \citep{ich07}, 
the stellar mass dependence of the X-ray properties of galaxies at $2<z<4$ 
\citep{yam09}, the evolution of the galaxy stellar mass function at $0.5<z<3.5$ 
\citep{kaj09}, the correlation between the stellar mass and surface brightness for 
galaxies at $0.3<z<3$ \citep{ich10}, the relation between the NIR morphology and   
star formation activity of galaxies at $0.8<z<1.2$ \citep{kon10}, 
the star formation activity as a function of stellar mass at $0.5<z<3.5$ \citep{kaj10}, 
and the evolution of quiescent galaxies as a function of stellar mass at $0.5<z<2.5$ 
\citep{kaj11}. 
\citet{yos10} also used these NIR images with NIR spectroscopic data 
of star-forming BzK galaxies at $z\sim2$ 
to investigate the star formation activity and stellar population of these galaxies. 
\begin{table*}
  \caption{Summary of the MODS observations}
\label{tab:first}
  \begin{center}
    \begin{tabular}{lccccc}
      \hline
             & center position &          & exposure time & FWHM (ch1, ch2) & 5$\sigma$ limit (ch1, ch2)\footnotemark[$*$]\\
      field & RA\hspace{1cm} Dec & band & (hour) & (arcsec) & (Vega mag)\\
      \hline
      GT-1 &  12:36:24.9  +62:10:43 & $J$ & 8.0 & 0.59\hspace{5mm} 0.59 & 24.3 \hspace{5mm} 24.2 \\
               &                                       & $H$ & 2.5 & 0.58\hspace{5mm} 0.59 & 23.3 \hspace{5mm} 23.1 \\ 
               &                                       & $K_{s}$ & 8.3 & 0.58\hspace{5mm} 0.53 & 23.1 \hspace{5mm} 23.2 \\
      \hline
      GT-2 & 12:36:47.8  +62:13:11 & $J$ & 28.2 & 0.48\hspace{5mm} 0.49 & 25.2 \hspace{5mm} 25.2 \\
               &    & $H$ & 5.7 & 0.46\hspace{5mm} 0.46 & 23.8 \hspace{5mm} 23.7 \\
               &    & $K_{s}$ & 28.0 & 0.45\hspace{5mm} 0.46 & 24.1 \hspace{5mm} 24.1 \\
      \hline
      GT-3 & 12:37:09.7  +62:15:58 & $J$ & 6.3 & 0.57\hspace{5mm} 0.58 & 24.4 \hspace{5mm} 24.3 \\
               &    & $H$ & 3.2 & 0.55\hspace{5mm} 0.55 & 23.1 \hspace{5mm} 23.1 \\
               &    & $K_{s}$ & 10.7 & 0.59\hspace{5mm} 0.60 & 23.2 \hspace{5mm} 23.2 \\
      \hline
      GT-4 & 12:37:31.8  +62:18:29 & $J$ & 9.1 & 0.58\hspace{5mm} 0.59 & 24.3 \hspace{5mm} 24.3 \\
               &    & $H$ & 4.3 & 0.58\hspace{5mm} 0.59 & 23.3 \hspace{5mm} 23.2 \\
               &    & $K_{s}$ & 9.8 & 0.59\hspace{5mm} 0.60 & 23.1 \hspace{5mm} 23.1 \\
      \hline
\multicolumn{6}{@{}l@{}}{\hbox to 0pt{\parbox{160mm}{\footnotesize \vspace{1mm}
\par\noindent
\footnotemark[$*$] the limiting magnitude is estimated from the background fluctuation 
measured with a aperture diameter of 2 $\times$ FWHM of the PSF. 
The total limiting magnitude for point sources,  
which is corrected for fluxes missed from the aperture, is $\sim$ 0.3 mag brighter for 
each field and band (see text for details). 
 }\hss}}
   \label{tab:field}
    \end{tabular}
  \end{center}
\end{table*}

In this paper, we present the NIR imaging data and 
a multi-wavelength photometric catalog of $K_{s}$-selected sources. 
Section 2 describes the observations. We give details of 
the data reduction in Section 3, and the properties of the reduced data in Section 4. 
In Section 5, we construct the source catalog and explain the catalog entries. 
Using the source catalog, we present the $K_{s}$-band number counts and distribution 
of NIR colors in Section 6. 
A summary is presented in Section 7. 

%We use a cosmology with H$_{\rm 0}$=70 km s$^{-1}$ Mpc$^{-1}$, 
%$\Omega_{\rm m}=0.3$ and $\Omega_{\rm \Lambda}=0.7$.
The Vega-referred magnitude system is used throughout this paper, 
unless stated otherwise.

%\noindent IMPORTANT NOTICE\\
%1. ``\verb|\draft|'' creates single column and double spaces format.\\
%2. If you comment out ``\verb|\draft|'', the output will be double column
%  and single space.\\
%3. For cross-references, the use of ``\verb|\label|, \verb|\ref|, \verb|\cite|'' 
%   and the thebibliography environment is strongly recommended. \\
%4. Do NOT use ``\verb|\def|, \verb|\renewcommand|''.\\
%5. Do NOT redefine commands provided by PASJ00.cls.\\

%\newpage

\section{Observation}
\label{sec:obs}
The deep 
$JHK_{s}$-band imaging observations of the GOODS-North were carried out with the 
Multi-Object InfraRed Camera and Spectrograph (MOIRCS; \cite{ich06};  
\cite{suz08}) on the Subaru Telescope. 
MOIRCS consists of two 2048 $\times$ 2048 HgCdTe HAWAII-2 detectors 
covering a field of view of 4 $\times$ 7 arcmin$^{2}$ 
 with a pixel scale of 0.117 arcsec/pixel. 
Our observations were made on 24 nights (including 8 half nights), in 
2006 April-May, 2007 March-May, and 2008 March-May. 
Figure \ref{fig:fov} shows the field layout of the MODS in the GOODS-North region.
Four MOIRCS pointings cover a field of $\sim$ 105 ($7 \times 15$) arcmin$^{2}$ 
(referred to as GT-1 to GT-4 field; see Figure \ref{fig:fov}). 
Nominal center positions are shown in Table \ref{tab:field}, and  adjacent fields 
overlap by $\sim$ 20 arcsec (Figure \ref{fig:fov}).
The position angle (PA) of the field of view was fixed to 45 degree for most of 
the observations. 
Since an engineering-grade chip, which is slightly less sensitive than science-grade one, 
 was used on one of the channels of the MOIRCS
(channel-1) in 2008 March-May, however, we also used the PA rotated by 180 degree
from the original value (i.e., PA $=$ 225 degree) 
equally in these observations 
to equalize the depths of the fields for the both channels.
For the $K_{s}$-band, we also used archival data of the MOIRCS obtained by other groups, 
namely, \citet{wan09} on 2005 December and 2006 June, and \citet{bun09} on 
2006 April. 

In the observations, we used the standard circular dithering pattern of the MOIRCS 
imaging mode (1 center + 8 surrounding pointings) 
with a diameter of $\sim 12$ arcsec. The center of 
the dithering pattern for each sequence was shifted randomly in a 10 arcsec box.
Exposure times for each frame 
 in the $J$, $H$, and $K_{s}$ bands are 100--260, 105--180, and 60--210 sec, respectively, 
which are split into 80--130, 20--40, and 20--70 sec sub-integrations, respectively, 
 to avoid the saturation of the detectors (the co-adding mode). 
The total integration time for each MOIRCS pointing is listed in Table \ref{tab:field},  
and the exposure maps for $J$, $H$, and $K_{s}$ bands are shown in Figure \ref{fig:fov}. 
For the GT-1, 3, and 4 fields, the integration times are 
6.3--9.1 hours in $J$ band, 2.5-4.3 hours in 
$H$ band and 8.3-10.7 hours in $K_{s}$ band, which were adjusted 
 so that the final depth of the survey data is homogeneous.
The GT-2 field,  which includes the Hubble Deep Field North (HDF-N, \cite{wil96}; 
see Figure \ref{fig:fov}), 
is  the ultra-deep field of the MODS, where the total integration times are longer 
than the other fields, namely, 28.2 hours in $J$ band, 5.7 hours in 
$H$ band and 28.0 hours in $K_{s}$ band.
\begin{figure*}
  \begin{center}
    \FigureFile(120mm,200mm){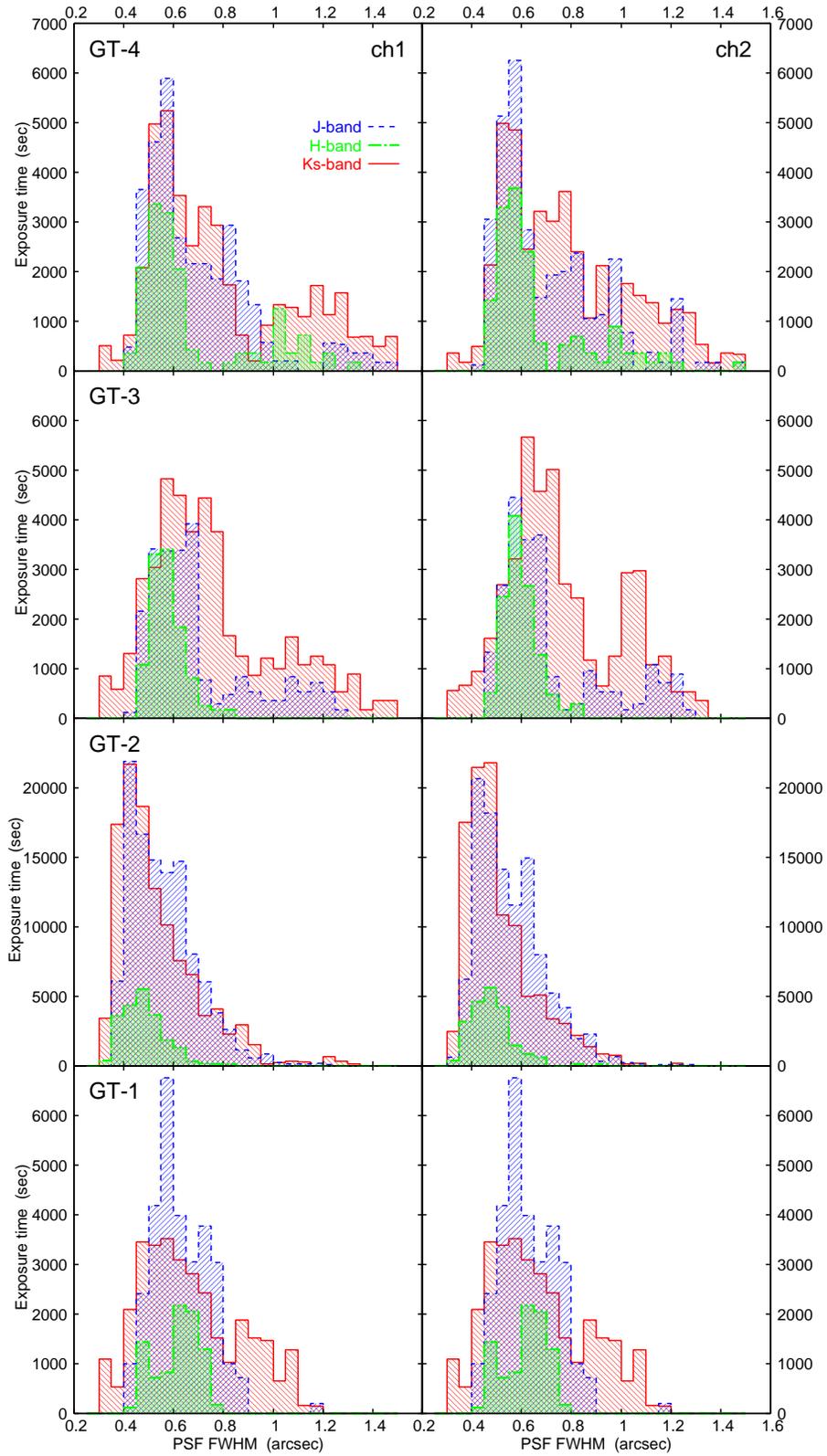}
    %%% \FigureFile(width,height){filename}
  \end{center}
\vspace{-3mm}
  \caption{Distribution of the seeing FWHM of the MODS observations
 for each field and chip.
The seeing was measured with relatively bright isolated point sources in each frame.
Blue, green and red histograms show those for $J$, $H$ and $K_{s}$-band data.
}
\label{fig:psf}
\end{figure*}

 Figure \ref{fig:psf} shows the seeing distribution of the data for each field. 
We measured FWHM of the point spread function (PSF) for  
each reduced (before combined) frame using relatively bright 
isolated point sources.  
Most of the data were obtained under relatively good seeing conditions ($\lesssim 
1.0$ arcsec). In excellent conditions, we preferentially observed the 
GT-2 field. 

A standard star P177-D \citep{leg06} was observed in a photometric night 
(1 April 2007) for flux calibration. 
For the $K_{s}$-band data,  we used the $K$-band magnitude of P177-D 
in \citet{leg06}  for the calibration.
%and use ``$K$-band'' (e.g., such as $K<23$) 
%for the magnitude measured in the MOIRCS $K_{s}$-band data in the following. 
The difference between $K_{s}$ and $K$-band magnitudes for P177-D is negligible. 
We obtained object frames for flux calibration immediately before the 
standard star observation at similar airmass in all 4 pointings in $JHK_{s}$ bands. 
The total integration time of these object frames for each field and each band is 
240--400 sec.

\section{Data Reduction}
The MOIRCS data were reduced basically using a 
purpose-made IRAF-based software package 
called {\it MCSRED}
\footnote{http://www.naoj.org/staff/ichi/MCSRED/mcsred\_e.html} \citep{tan10}. 
The reduction was performed independently for each chip, and then the combined 
images for the different chips (channel-1 and 2) 
and fields (GT-1 to GT-4) were mosaiced finally. 
The basic procedure in the MCSRED is 1) making a quick object mask for each raw 
frame, 2) making a self-sky flat frame by median stacking of masked raw frames, 
3) flat-fielding, 4) making a median sky frame for each flat-fielded frame from 
6-8 frames before and after the frame, 
5) performing sky subtraction, 
6) removing residual sky gradients by fitting each quadrant of the object-masked 
frame with a 3rd-order polynomial surface and subtracting it from the unmasked image,  
7) correcting the optical distortion, 
8) measuring positional offsets among the frames and aligning, 
9) stacking the frames. 
Although the self-sky flat for NIR cameras often suffers from thermal emission 
from the camera, MOIRCS was designed so that the thermal emission from the camera and
telescope was minimum \citep{suz08}. 
We confirmed that the self-sky flat agrees within 5\% with a dome-flat made from  
the ``on $-$ off'' frame even in $K_{s}$ band, where the effect of thermal emission 
is expected to be maximum. 
We also checked the photometry at the different positions in the detectors using 
objects in the overlap regions between the different chips and fields (see below).   
Once the reduction sequence was finished, 
we used the combined image to make a more accurate object 
mask which includes fainer objects than a quick mask, and then repeated the procedures
 of 2)--9). Such ``second-pass'' procedure allows us to remove the effect of faint objects 
on the flat-fielding and sky subtraction. 
\begin{figure*}
  \begin{center}
    \FigureFile(150mm,100mm){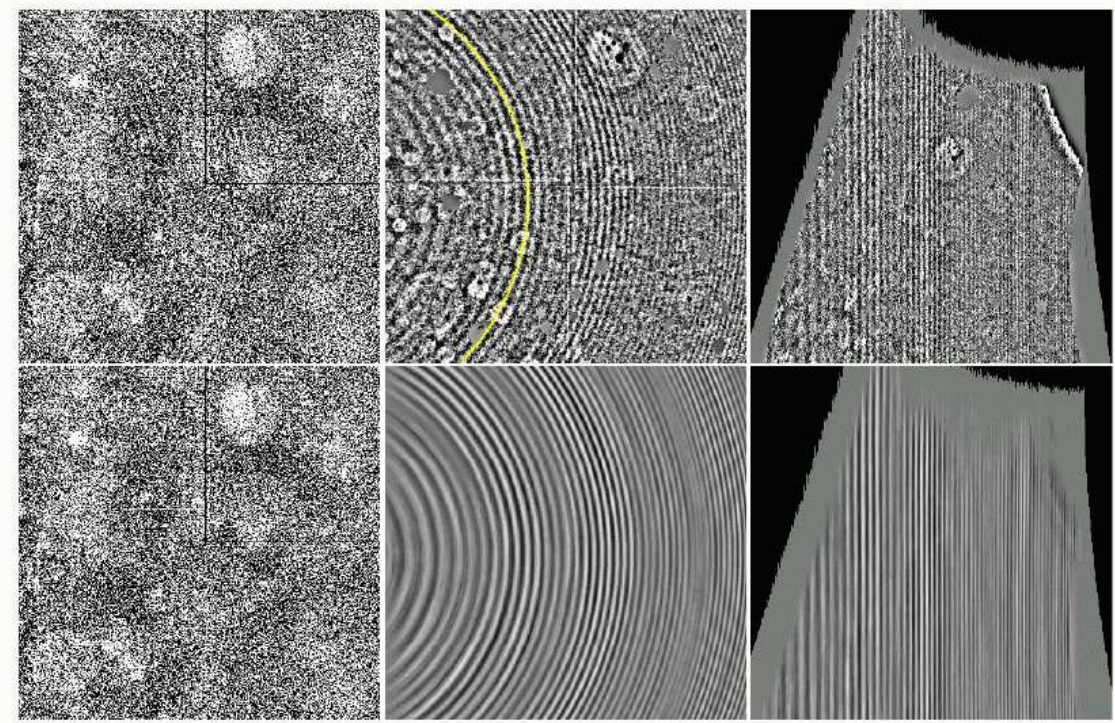}
    %%% \FigureFile(width,height){filename}
  \end{center}
\vspace{-3mm}
  \caption{Example of the defringing process for a $K_{s}$-band raw 
frame of channel-2. 
{\bf top-left:} a raw frame is quick flat-fielded by a dome-flat and sky-subtracted. 
{\bf top-middle:} objects in the frame is masked and then the frame is convolved 
with a Gaussian kernel to reduce noise.
{\bf top-right:} the frame is converted from 
 Cartesian coordinate system ($x$, $y$)
to the polar coordinate system ($r$, $\theta$).
{\bf bottom-right:} the converted image is smoothed along the vertical direction 
with a median filter of 1 $\times$ 401 pixels.
{\bf bottom-middle:} the frame is converted back into 
the Cartesian coordinate system to make a fringe pattern frame.
{\bf bottom-left:} the fringe pattern frame is subtracted from the original frame. 
}
\label{fig:fringe}
\end{figure*}

In addition to the basic procedure of the {\it MCSRED}, we also performed 
a defringing process. 
Each raw frame obtained in the MODS, which was carried out before the 
installation of the ``fringe-free'' type filter,  contains a circular fringe pattern 
produced by the interference of OH air glows. 
Since the amplitude and center position of the circular pattern varied with time, 
the fringe pattern cannot be completely removed by the standard sky subtraction 
procedure. These residual patterns 
could systematically affect the source detection, photometry, or morphological 
analysis for objects in the combined images. 
%prevent the combined image from reaching a depth 
%expected from the long exposure time of the MODS. 
Figure \ref{fig:fringe} shows an example of the defringing process for a raw 
frame (channel-2, $K_{s}$-band). 
The circular fringe pattern can be seen in  a frame which is flat-fielded by 
a dome-flat and is then sky-subtracted with the 3rd-order polynomial surface fitting 
(top-left panel of the figure).
In order to reduce noise and 
highlight the fringe pattern, we masked objects in the image and 
convolved the frame with a Gaussian kernel with $\sigma = 2$ pixel (top-middle panel).  
Then we determined a center of the circular fringe pattern (e.g., 
yellow curve in the top-middle panel), and converted the image in the 
Cartesian coordinate system ($x$, $y$)
to that in the polar coordinate system ($r$, $\theta$), where the fringe pattern 
becomes vertical stripes (top-right panel).  
The converted image was smoothed along the vertical direction (azimuthal 
direction in the original Cartesian coordinate) 
with a median filter of 1 $\times$ 401 pixels to reduce noise and fill in the 
masked region (bottom-right panel). Then the smoothed image was converted back into 
the Cartesian coordinate system and was multiplied 
by the dome-flat to obtain a fringe frame for the original raw data (bottom-middle panel).
Finally, the fringe frame was subtracted from the raw frame (bottom-left panel).

We used this defringing process twice in the reduction procedure.
First, we applied the defringing process to raw data and used 
the defringed frames to make a sky-flat frame which is not affected by the 
fringe pattern. With this sky-flat frame, we performed flat-fielding of original 
(not-defringed) raw data, and then 
followed the standard sky-subtraction procedures mentioned above. 
After the sky subtraction, we applied the same 
defringing process to the sky-subtracted frames to remove the residual of 
the time-variant fringe pattern. 
The defringing process after the sky-subtraction has the advantage of much  
smaller amplitudes of the residual patterns than in a case where the process is 
directly applied to raw data. 

Before the stacking procedure, all object frames were scaled to the count level of 
the reference frame for the flux calibration 
mentioned in the previous section, using relatively 
bright unsaturated sources. 
We discarded the frames whose count level was less than 70\% of the 
calibration frame, 
because 
we often failed in the sky subtraction of these frames 
%the sky subtraction of these frames often terribly 
%failed 
due to the bad and unstable sky condition.
In the stacking, we used the weighted-average with 3$\sigma$ clipping.  
In order to maximize the signal-to-noise (S/N) ratio for the surface brightness of 
the combined image, 
each frame was weighted by 1/$\sigma_{\rm pix}^{2}$, where $\sigma_{\rm pix}$ 
is the standard deviation in the pixel-to-pixel statistics of the frame whose count level was 
scaled to the reference frame. 
We did not weight the images with seeing size. 
Instead, we made several combined images for each field and each band 
using the data sets grouped in different seeing sizes, 
%with various seeing limits, 
and measured the background fluctuation of the combined images with apertures 
of various sizes. Based on the results, we decided to discard 
frames with a seeing size of FWHM $> 1.2 $ arcsec for the GT-1, 3, and 4 fields and 
$> 0.8 $ arcsec for the GT-2 field from the stacking. 
For $J$ and $K_{s}$-band data in the GT-2 field, we also provided the high-resolution 
combined image only from the data sets with a seeing size of $<0.5$ arcsec 
to investigate the morphology of galaxies in the NIR wavelength \citep{kon10}.
In addition to the combined image, we also provided the exposure map and 
RMS map which gives an estimate of the noise level at each pixel. 
In order to make the RMS map, we first divided the ``sigma map'' 
outputted from the IRAF/IMCOMBINE task by the square root image of 
the exposure map, and 
then normalized the resulting image so that the average of pixels in the RMS map 
corresponds to the standard deviation in the pixel-to-pixel statistics of the combined 
object image. 

In order to mosaic the combined images for the different chips and fields, at first, 
each combined image was rescaled to a common 
zero point of 26.0 AB mag (for all $JHK_{s}$ bands). 
We then determined 
the position of each combined image via astrometry 
%the astrometry of each combined image 
using the 
GOODS HST/ACS version 2.0 data, whose world coordinates agree well 
with catalogs from the SDSS, 2MASS, and deep VLA 20cm data 
(\cite{gia04}; M. Giavalisco and the GOODS Team, 2010, in preparation). 
Relatively bright isolated stars and galaxies with early-type morphology in each 
combined image were cross-matched with those in the HST/ACS 
$z_{\rm F850LP}$-band image. About 100 such objects 
were used  for  each combined image 
and  the world coordinates were determined 
with a root mean square scatter 
of $\sim $ 0.03 -- 0.05 arcsec, which corresponds to $\sim $ 0.25 -- 0.45 pixels.
Based on the determined world coordinates, the combined images were 
transformed into a common grid of the expanded image of the 
GT-2 channel-1 data. 
We performed the photometry of objects in the overlap regions between the different 
chips and fields to check the zeropoint over the detectors in each band. 
The magnitudes at the different positions in the detectors agree well and 
no significant systematic difference is seen. The typical measured average
difference and root mean square are 
$0.03 \pm 0.10$ mag in $J$ band, $0.02 \pm 0.19$ mag in $H$ band, and 
$0.01 \pm 0.11$ mag in $K_{s}$ band.
In the mosaicing process, the pixels in the overlap regions of the different images 
were combined with the weighted-average without sigma clipping. 
The inverse square of the RMS map for each combined image was used as the 
weight.

\section{Reduced  images}
\subsection{Area coverage}
\begin{figure*}
  \begin{center}
    \FigureFile(105mm,170mm){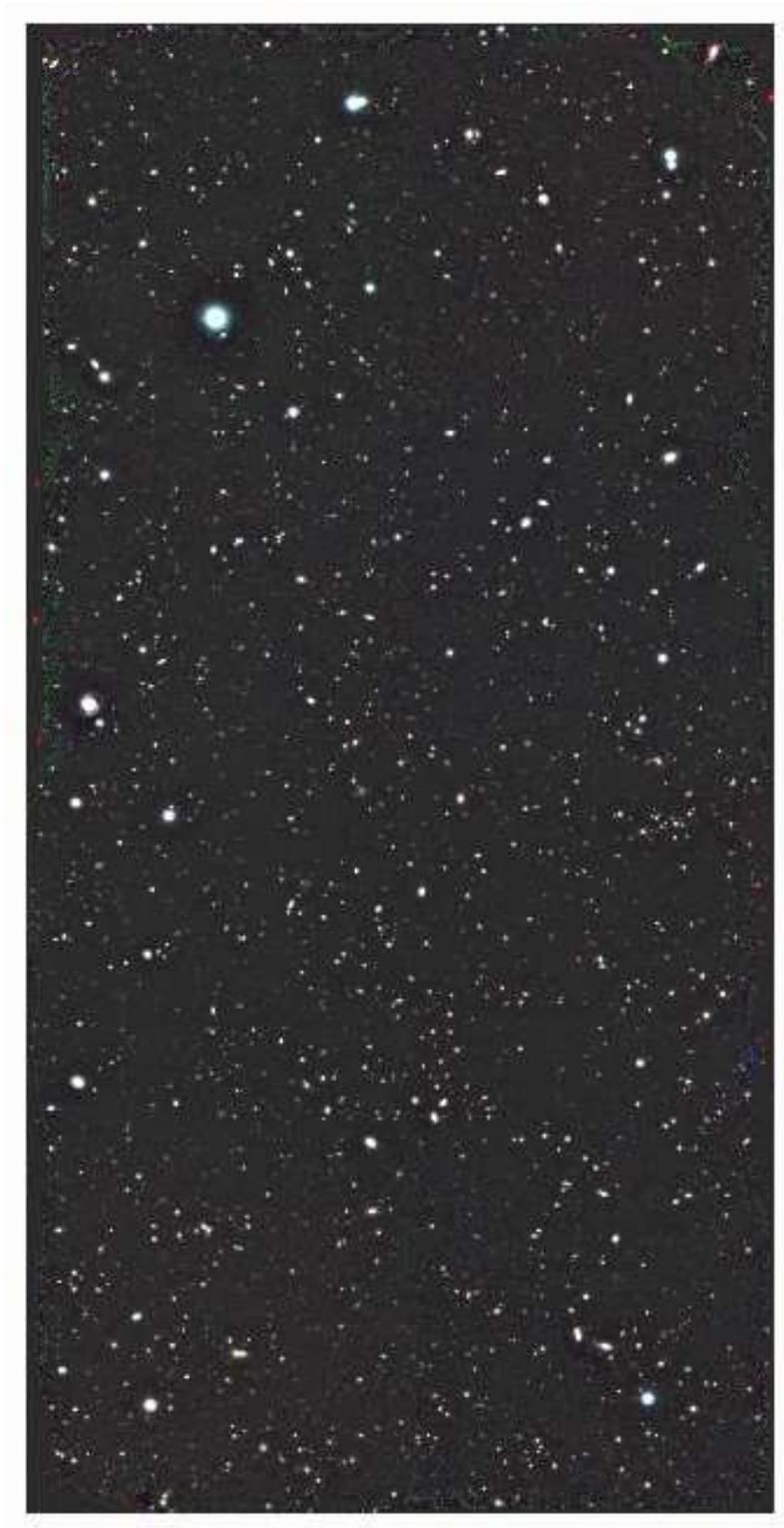}
    %%% \FigureFile(width,height){filename}
  \end{center}
  \caption{$JHK_{s}$-band pseudo-color image of the MODS}\label{fig:modsjhk}
\end{figure*}
The reduced $J$, $H$, $K_{s}$-band mosaic images have a pixel scale of 0.117 
arcsec and a image size of 4200 $\times$ 8400 pixels including the margin. 
The total exposure times t$_{\rm exp}$ 
for the images are shorter toward the outer edges as the  
result of the dithering technique in the observations mentioned above. 
The area in the $J$, $H$, and $K_{s}$-band images with t$_{\rm exp} > $ 3600 (7200) sec 
is 106.3 (104.3), 103.3 (98.5), and 118.6 (114.1) arcmin$^{2}$, 
respectively. The larger area coverage in $K_{s}$ band reflects the contribution from 
the observations by other groups mentioned above, 
whose field configurations are different from ours. 
If we use the homogeneous depth in the central region of each field as a baseline, 
 the area with $\sigma < 2 \times \sigma_{\rm (central \ region)}$ in the RMS maps is 
104.6, 104.8, and 109.8 arcmin$^{2}$ for the $J$, $H$, and $K_{s}$-band images. 
In order to make a $K_{s}$-selected source catalog in the next section, we use the 
area with t$_{\rm exp} > 3600 $ sec in the $H$-band image, which ensures 
 $\sigma < 2 \times \sigma_{\rm (central \ region)}$ in all three bands.
The area with t$_{\rm exp} > 3600 $ sec in $H$ band 
is 103.3 arcmin$^{2}$ for the whole field and 28.2 arcmin$^{2}$ for the deep GT-2 field. 

\subsection{Image quality and PSF matching}
The PSF of the combined image in each chip and field was measured using 
relatively bright isolated point sources.  
The FWHMs of the PSF are listed in Table \ref{tab:field}.
The PSF of the $J$, $H$, and $K_{s}$-band images for the GT-1, 3, and 4 fields 
ranges 0.53 -- 0.60 arcsec (0.58--0.59 arcsec for the most images).  
The images for the GT-2 field, which were obtained preferentially in the excellent 
 condition, have sharper PSF with FWHMs of 0.45 -- 0.49 arcsec.

Since the different field images have different PSF FWHMs,  
 there is PSF variance in each mosaic image. 
In order to measure colors of objects with consistent apertures across all bands, 
we made PSF-matched mosaic images by convolving all images to a common PSF, 
namely that of the GT-3 $K_{s}$-band image, which has the poorest resolution 
(FWHM $=$ 0.60 arcsec), before the final mosaicing. 
For the GT-3 $J$-band and GT-4 $K_{s}$-band data, we used the combined images 
provided from the frames with  PSF FWHMs of $< 1.0$ arcsec for the PSF-matched 
mosaic images  
%instead of $< 1.2$ arcsec for the original mosaic images, 
in order to secure a 
resolution for the proper matching of the PSF to that of the GT-3 $K_{s}$-band image.
Figure \ref{fig:modsjhk} shows a $JHK_{s}$-band pseudo-color image made from 
the PSF-matched mosaic images.
We also made PSF-matched mosaic images only for the GT-2 field 
with a higher resolution. These images were provided from the combined images which 
were convolved to the resolution of the GT-2 $J$-band image (FWHM $=$ 0.5 arcsec).  
In Section \ref{sec:color}, we compare the growth curve of the point sources 
in the PSF-matched images among the different bands 
in order to check the accuracy of the PSF-matching.

\subsection{Background noise and limiting magnitude}
\begin{figure*}
  \begin{center}
    \FigureFile(170mm,120mm){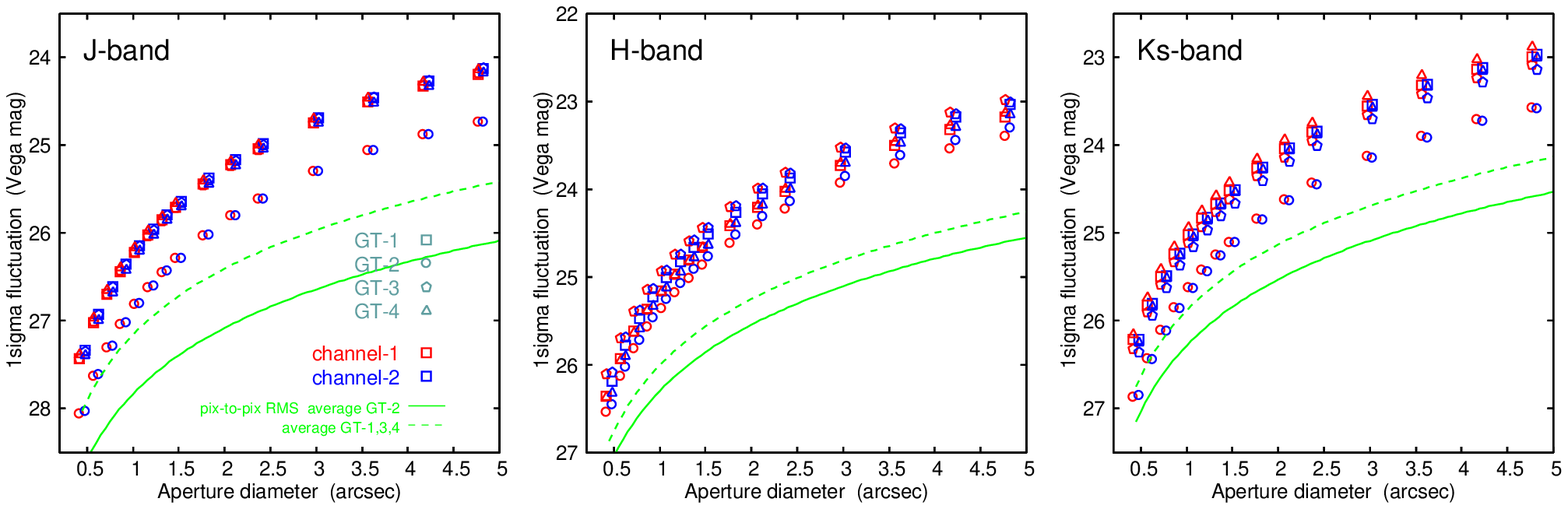}
    %%% \FigureFile(width,height){filename}
  \end{center}
\vspace{-3mm}
  \caption{Background sky fluctuation as a function of aperture diameter for each band. 
Different symbols represent the different fields. Red and blue symbols show 
those for channel-1 and 2, respectively.
Dashed and solid lines show 
the fluctuation expected from the pixel-to-pixel root mean square scaled by 
the square root of the aperture area under the assumption of the Poisson statistics 
for the GT-1, 3, and 4 fields and the GT-2 field, respectively.  
}
\label{fig:noise}
\end{figure*}

In order to estimate the background noise and derive the limiting magnitude, 
we directly 
measured fluxes of circular apertures placed at random positions 
on the combined images where 
objects were masked and replaced with pseudo noise images.
In order to provide the pseudo noise images, we averaged 10 randomly-shifted 
combined images (where objects were masked) and multiplied by the square 
root of the number of the frames used for the average in each pixel, assuming 
the Poisson statistics. 
We then calculated the root mean square of these measurements, and 
 repeated the measurements with various aperture diameters.
Figure \ref{fig:noise} shows the results for the different chips and fields at each 
band. For comparison, the expected values from the pixel-to-pixel root mean 
square scaled by 
the square root of the aperture area, which is calculated under the assumption that 
the pixel-to-pixel noise is the Poisson distribution and uncorrelated each other, 
are also shown in the figure (solid and dashed curves). 
The empirically measured background noise is larger than that expected from the 
scaled pixel-to-pixel noise,  because adjacent pixels in the combined images 
become correlated through the reduction procedures such as the distortion correction 
or the registration of the frames. 
%The small effect of (masked or unmasked) 
%objects on the median sky frame could also cause additional variance 
%in the combined images. 

We estimated 5$\sigma$ limiting magnitude for each chip and field 
using the background fluctuation measured with a aperture diameter of 2 $\times$ 
FWHM of the PSF. These limiting magnitudes are listed in Table \ref{tab:field}.
For GT-1, 3, and 4 fields, the depth of the data is relatively homogeneous and the 
limiting magnitudes are $J \sim 24.2$, $H \sim 23.1$, and $K_{s} \sim 23.1$. 
In the GT-2 field, the limiting magnitudes are $J \sim 25.1$, $H \sim 23.7$, and 
$K_{s} \sim 24.1$, which are $\sim$ 1 magnitude deeper than the GT-1, 3, and 4 fields.
In order to derive the total limiting magnitude for point sources, we need to estimate 
the aperture correction. 
In Appendix \ref{sec:apcor}, we carry out simulations with the IRAF/ARTDATA package 
and find that the aperture magnitude with a diameter of 2 $\times$ FWHM of the 
PSF is fainter by $\sim$ 0.28--0.29 mag than the total magnitude for point sources.
The total limiting magnitudes for point sources corrected for the missed fluxes are also 
listed in Table \ref{tab:field}.

These limiting magnitudes show that the MODS data are one of the deepest NIR 
imaging data over a area of $\sim$ 30 to $\sim$ 100 arcmin$^{2}$. 
In particular, the $K_{s}$-band data in the GT-2 field is the deepest image obtained 
to date at the wavelength.   
For example, the ESO GOODS-South NIR imaging survey with the VLT/ISAAC 
achieved 5$\sigma$ limiting magnitudes of $J \sim 24.0$, $H \sim 23.1$, and 
$K_{s} \sim 22.5$ over $\sim 130$ arcmin$^{2}$, and the ultra deep $K_{s}$-band data with a 
5$\sigma$ limit of $K_{s} \sim 23.7$ was also obtained over $\sim$ 6.7 
arcmin$^{2}$ in the HUDF \citep{ret10}. 
The NIR imaging data obtained in the FIRES survey reach 5$\sigma$ limits of 
$J \sim 25.3$, $H \sim 24.2$, and $K_{s} \sim 23.8$ over $\sim$ 4.7 arcmin$^{2}$, and 
$J \sim 24.6$, $H \sim 23.7$, and $K_{s} \sim 23.1$ over $\sim$ 26.2 arcmin$^{2}$ 
(\cite{lab03}; \cite{for06}). \citet{wan10} performed a wider but shallower $K_{s}$-band  
imaging survey than the MODS 
in the GOODS-North region with the CFHT/WIRCam 
and achieved a 5$\sigma$ limit of $K_{s} \sim 22.7$ over $\sim$ 200 arcmin$^{2}$.

\section{$K_{s}$-selected catalog}
We constructed multi-band photometric catalogs of $K_{s}$-selected sources 
in the MODS field, 
using the MODS data and multi-wavelength public data of the GOODS
survey, KPNO/MOSAIC $U$-band data \citep{cap04}, HST/ACS $B$, $V$, $i$, 
and $z$-band version 2.0 data (Giavalisco et al., in preparation; \cite{gia04}), 
and Spitzer/IRAC 3.6, 4.5, 5.8, and 
8.0  $\mu$m-band DR1 and DR2 data (M. Dickinson et al., in preparation). 
Two catalogs are provided; 
one includes objects in the whole MODS field (``wide'' catalog), 
and the other includes only those in the GT-2 field (``deep'' catalog). 
For the deep catalog, the multi-band photometry was performed with a slightly 
smaller aperture (1.0 arcsec diameter) 
in the sharper PSF-matched images with FWHM of 
$\sim $ 0.5 arcsec than that for the wide catalog (1.2 arcsec diameter). 
It should be noted that the source detection was performed in the  
$K_{s}$-band non-convolved (not PSF-matched) image in both cases. 
We describe a method of constructing the catalogs and the catalog entries 
in the following. 

\subsection{Source detection and completeness}
\label{sec:detect}
The source detection was performed in the $K_{s}$-band image 
using the SExtractor software version 2.5.0 \citep{ber96}.
A detection threshold of 1.3 times the local background root mean square
 over 12 connected 
pixels was used. In order to take account of variance of the background noise 
among different positions 
in the image, we used the RMS map as a weight image in the source detection 
procedure. 
The ``local''-type background estimate of the SExtractor was used for the 
photometry. 
We adopted MAG\_AUTO from the SExtractor as the total $K_{s}$-band magnitudes 
of the detected objects. 
In Appendix \ref{sec:apcor}, we performed simulations with the IRAF/ARTDATA 
package to estimate fluxes missed from the Kron apertures of the MAG\_AUTO 
for point sources as a function of $K_{s}$-band magnitude. 
According to the results of the simulations, 
the missed fluxes for point sources 
are expected to be $\sim$ 2--3\% at $K\sim$ 17--18 mag and 
become $\sim$ 10--15\% near the 5$\sigma$ limiting magnitudes on average. 
We also list the magnitude corrected for these average missed fluxes 
(MAG\_AUTO\_COR) in the catalogs as well as the original MAG\_AUTO. 
The photometric error for MAG\_AUTO was estimated 
from the background fluctuation shown in Figure \ref{fig:noise} 
measured with a circular aperture whose area is 
the same as the elliptical Kron aperture used in MAG\_AUTO for each object. 
Since the background noise shown in Figure \ref{fig:noise} is averaged over 
each chip, we also used the RMS map at the object position 
to take account of the noise variation in each chip and estimate the background noise 
for the object.  Since the RMS map includes an effect of the signal from the objects, 
we did not add the photon noise from the objects to the photometric error separately. 
\begin{figure*}
  \begin{center}
    \FigureFile(115mm,190mm){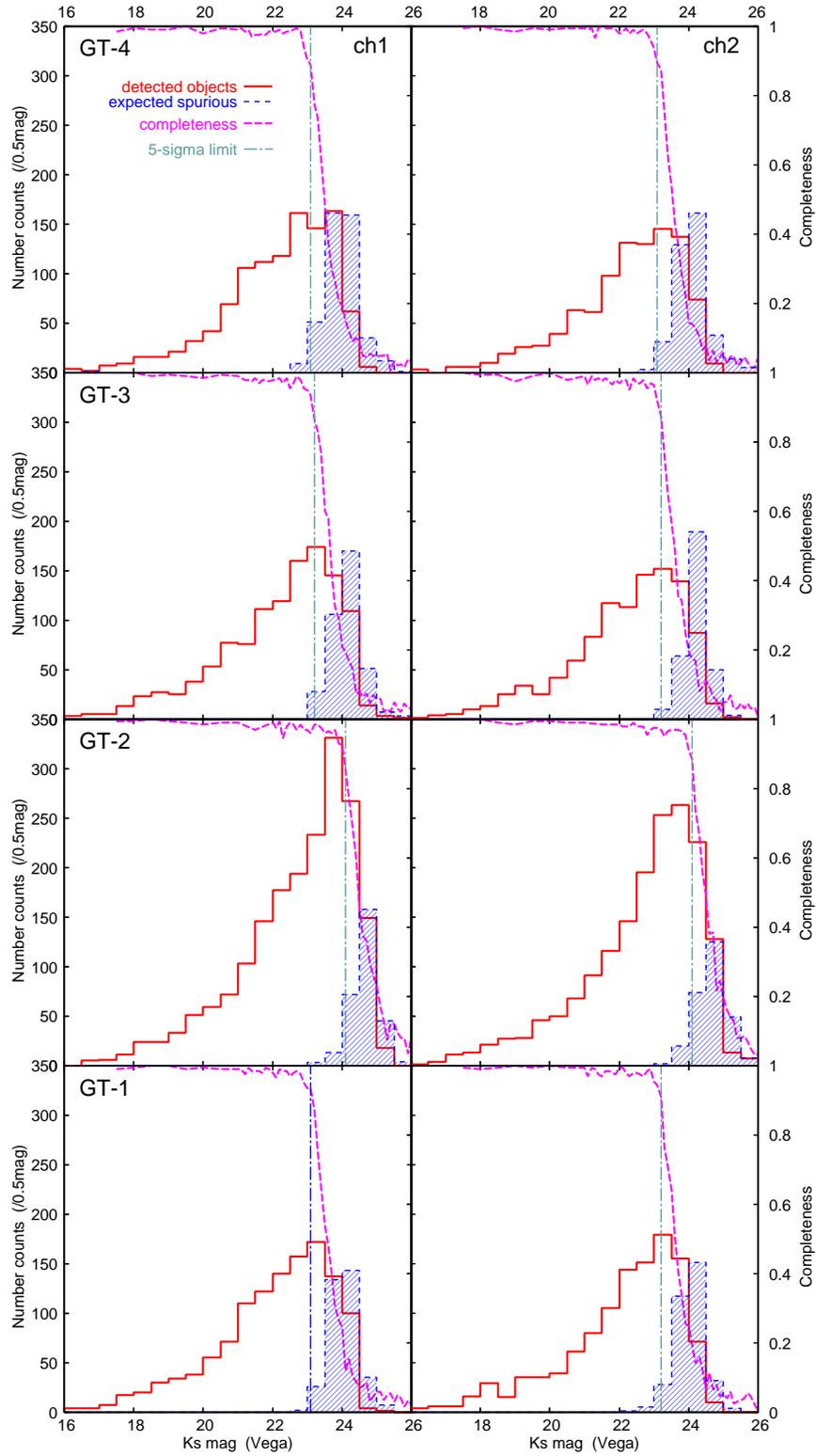}
    %%% \FigureFile(width,height){filename}
  \end{center}
\vspace{-3mm}
  \caption{$K_{s}$-band number counts of detected objects (open histogram)
and expected false detections (shaded histogram) for each field and chip.
Vertical dashed-dotted line in each panel shows the 5$\sigma$ limiting magnitude.
Dashed line shows the detection completeness for the point sources.
}\label{fig:comp}
\end{figure*}

As mentioned in the previous section, 
we included only objects with  t$_{\rm exp} > 3600 $ sec in the $H$ band into the catalogs 
to ensure relatively uniform depth in $JHK_{s}$ bands.
Total 11660 objects were detected in the selected area of 103.3 arcmin$^{2}$. 
In order to examine the fraction of false detections, we performed a simulation of 
the source detection on the inverse $K_{s}$-band image with the exact same 
detection parameters. Figure \ref{fig:comp} shows number counts of the objects
detected in the original image 
and the false detections expected from the simulation for each chip and field.
The 5$\sigma$ limiting magnitude estimated from the background fluctuation 
in the previous section is also shown in the figure.
The expected false detection rate increases at faint magnitude rapidly, but it 
 is negligible ($\lesssim 1 $ \%) at brighter than the 5$\sigma$ limiting magnitude. 
Below the limiting magnitude ($\lesssim$ 1-2$\sigma$ of the background noise), 
the numbers of the detections in the inverse image 
tend to be larger than those in the original image.
This may be because the noise property at very faint level is affected by the 
reduction procedures such as the median sky subtraction.  

Around very bright stars, there are also spurious sources caused by 
the image persistence in the detectors of MOIRCS. 
These spurious features are relatively faint 
and appear at $\sim$ 8--9 arcsec from the bright stars, 
which corresponds to the dither length between successive positions 
used in the observations (Section \ref{sec:obs}). 
Although we identified 15 spurious sources around very bright 
stars in the $K_{s}$-band image and excluded them from the catalogs, there could remain  
such spurious sources in the catalogs. 
Objects near very bright stars should be handled with caution.
 
We also estimated the detection completeness of our $K_{s}$-selected sample 
using Monte Carlo simulations with the IRAF/ARTDATA package. 
The detection completeness 
actually depends on many parameters such as magnitude, size, surface brightness
profile, crowdedness, and so on, in a complex way. 
We here examine the completeness for the point source as a guide. 
In the simulations, a point source with the corresponding PSF was added at random
position in the image. We then ran the SExtractor with the same detection parameters 
and check whether the added object was detected or not. 
For each chip and field, 200 such simulations were done for a given magnitude of the 
point source. The results are shown in Figure \ref{fig:comp}. 
The detection completeness for the point sources is $\sim$ 85--90 \% level at the 
5$\sigma$ limiting magnitude. The completeness decreases rapidly below the 
limiting magnitude, and becomes $\sim$ 50 \% at  0.5 mag below the limit.

\subsection{Color measurements}
\label{sec:color}
We measured the optical-to-MIR SEDs of the detected objects, 
using  the $U$ (KPNO/MOSAIC), $BViz$ (HST/ACS), $JHK_{s}$ (MOIRCS), 3.6, 4.5, 5.8, 
8.0 $\mu$m (Spitzer/IRAC) -band images.
These multi-band images were aligned to the $K_{s}$-band mosaic image. 
We convolved the ACS images with the IRAF/PSFMATCH task  
to match the PSF-matched images of the MODS. 
\begin{figure*}
  \begin{center}
    \FigureFile(150mm,120mm){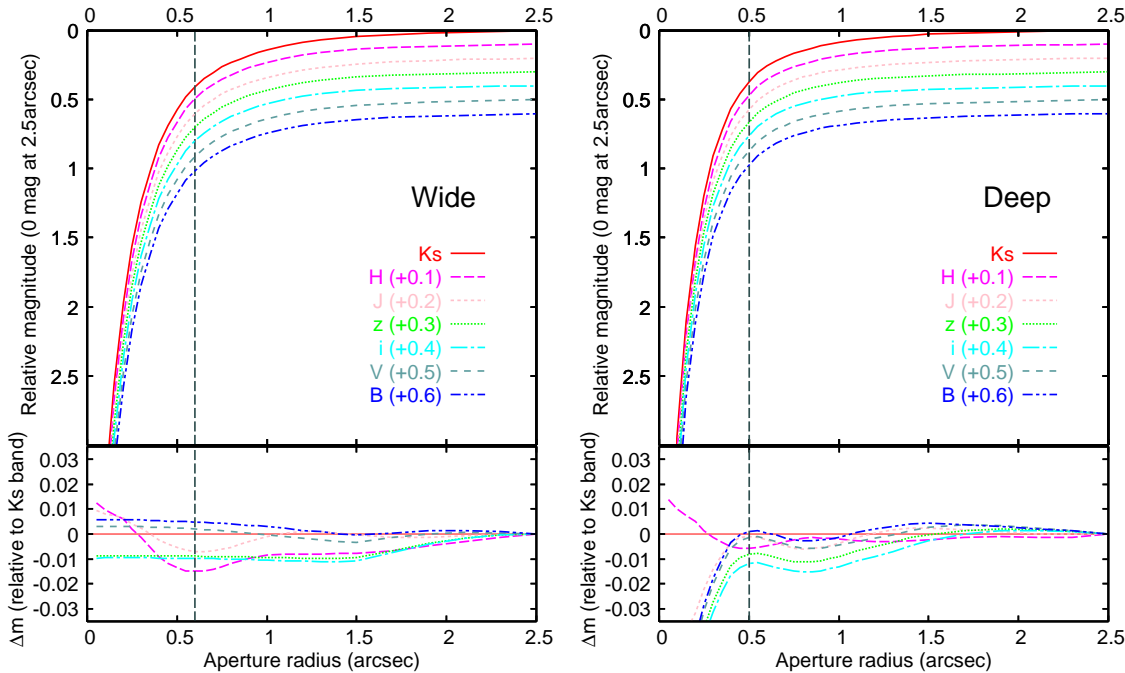}
    %%% \FigureFile(width,height){filename}
  \end{center}
\vspace{-3mm}
  \caption{Comparisons of the growth curves for point sources in the PSF-matched 
images among the different bands ($BVizJHK_{s}$) for the wide (left) and deep (right) 
fields. The growth curves are normalized to 0 mag at 2.5 arcsec radius. 
The growth curves in the top panels are plotted with vertical offsets (by 0.1 mag) 
for clarity. The bottom panels show the differences from the $K_{s}$ band 
in the growth curves. Vertical long-dashed lines represent the aperture radius used 
for the color measurements in the wide and deep catalogs.   
}
\label{fig:growth}
\end{figure*}
Figure \ref{fig:growth} compares the growth curves of the point sources in the 
PSF-matched images among the different bands (ACS and MOIRCS bands). 
For the comparison, we made an average PSF from $\sim$ 30 
relatively bright, isolated point sources in  each PSF-matched image.
The growth curves of the stars in the different bands agree within 
$\pm$ 0.02 mag at radii larger than 0.5 arcsec.  

For the color measurements, we basically used a fixed aperture diameter of 1.2 
arcsec for the wide catalog and 1.0 arcsec for the deep catalog, respectively, as 
mentioned above.
For MOSAIC ($U$-band) and IRAC (3.6, 4.5, 5.8, and 8.0 $\mu$m-band) data, 
whose spatial resolution is much poorer than that of the ACS and MOIRCS data, 
we first performed the aperture photometry for the wide (deep) catalog with 
aperture sizes of 2.28 (2.39), 2.89 (3.07), 3.05 (3.24), 3.80 (4.02), and 4.10 (4.33) arcsec, 
respectively, and then applied the aperture correction by using  the light profiles 
of the $B$- and $K_{s}$-band images convolved to match the resolution of 
$U$-band and IRAC images, respectively.
These aperture sizes for the $U$-band and IRAC data were selected so that 
the fraction of the flux sampled by the aperture was the same 
 as  that in the ACS and MOIRCS images for point sources. 
The SExtractor dual-image mode was used for the photometry. 

Before running SExtractor, we also performed the subtraction of 
the contribution from neighboring sources for objects in the crowded region.   
For the ACS and MOIRCS data, we first picked up all sources in the non-convolved 
images within 12 arcsec from the $K_{s}$-selected object in question. 
We checked  
 whether the pixel on the $K_{s}$-band image at the center position of each picked source  
belongs to the object in question or not in order to judge if it is the corresponding source. 
If a picked source did not correspond to the $K_{s}$-selected object, this source was 
extracted from the non-convolved image and then was convolved and subtracted from 
the PSF-matched (convolved) image. For the $U$-band and IRAC data, 
we used the $B$- and $K_{s}$-band images for the subtraction of the contribution from 
neighboring sources. In this case, all sources picked in the non-convolved $B$- or 
$K_{s}$-band image were extracted and convolved to match the $U$-band  or IRAC 
data at first. 
Then we measured a flux of the central region of the convolved image for each picked 
source. 
Here, the central region is 
defined by the pixels belonging to the source in the non-convolved image.  
We also measured a flux 
 of the corresponding (central) region in the $U$-band or IRAC image, 
and scaled the flux of 
the convolved $B$- or $K_{s}$-band image to match that in the $U$-band or IRAC image. 
This flux scaling was done in order of the brightness of the sources
in the $U$-band or IRAC image. 
Once the scaling for the brightest source in a crowded region 
was done, the scaled image was subtracted 
from the image of the region and then the scaling for the second brightest source was 
done. After the scaling for all sources finished, the scaled images for sources which 
do not correspond to the object in question were subtracted from the original 
$U$-band or IRAC image. 

The error for the aperture photometry was estimated in a similar way 
with that for MAG\_AUTO mentioned above. 
We  measured the background fluctuation for the convolved ACS and MOIRCS data  
and the MOSAIC and IRAC data with the aperture used in each band, 
and similarly used a RMS map for each band 
to estimate the background noise at each object position in the image. 
When the measured flux for an object is less than 2$\sigma$ of the background 
noise, the upper limit is assigned.
For objects in crowded regions, we also performed the aperture photometry without 
the subtraction of the contribution from neighboring sources, and conservatively 
added a half of the difference between the results with and without the subtraction 
process to the photometric error. 

We included only objects which are detected above 2$\sigma$ level in more than 
three bands ($K_{s}$ band and other two bands) in the main catalogs, because 
it is difficult to estimate the photometric redshift of those detected only in 
one or two bands.   Of objects detected in the $K_{s}$ band above 2$\sigma$ level, 
933/10806 and 189/3983 were excluded from the wide and deep catalogs by this 
criterion. We provide complementary catalogs of these sources separately.  
Most of these excluded objects are fainter than the 5$\sigma$ limiting magnitude 
of the $K_{s}$-band data.

\subsection{Redshifts}
\label{sec:redshift}
\begin{figure*}
  \begin{center}
    \FigureFile(120mm,180mm){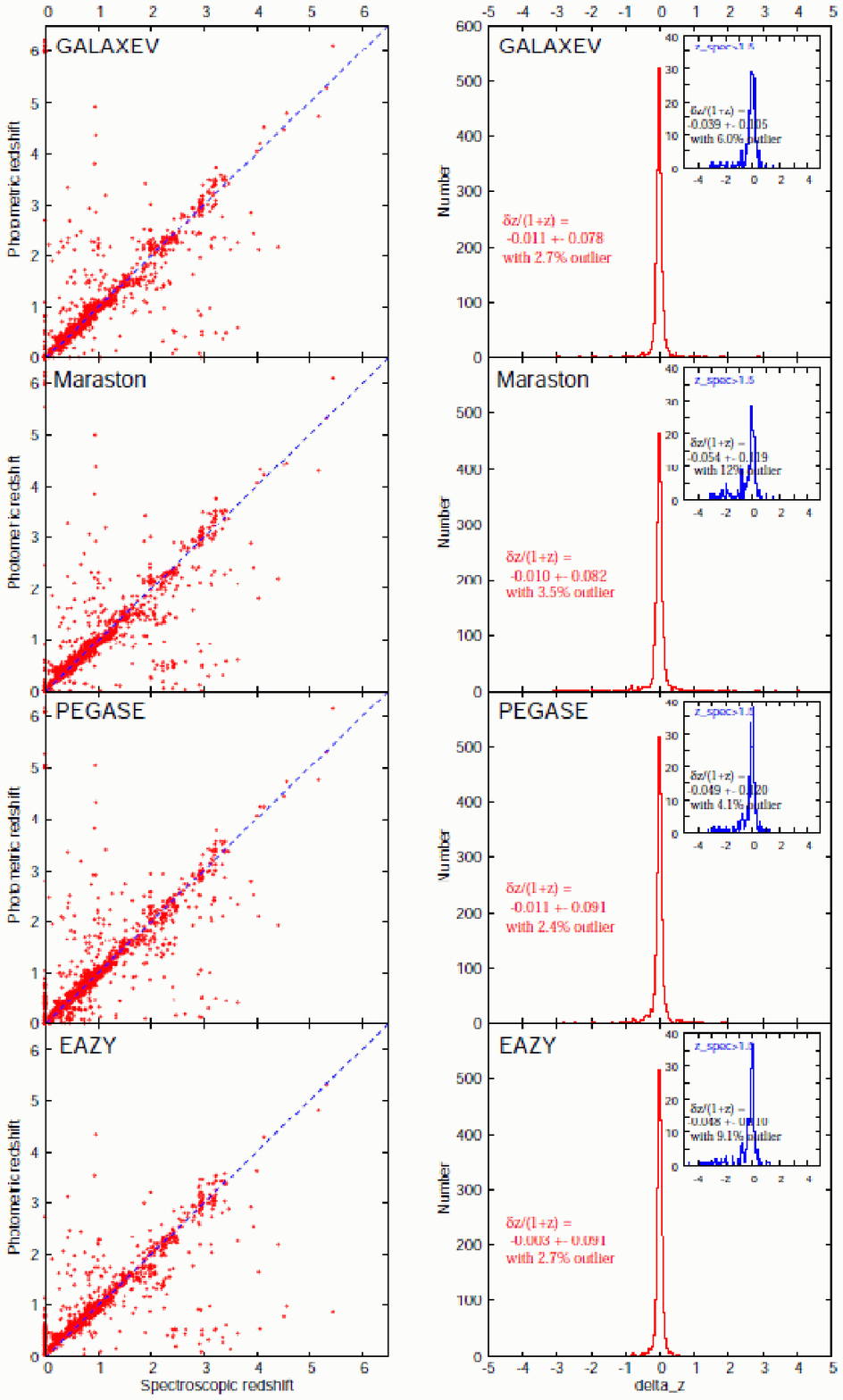}
    %%% \FigureFile(width,height){filename}
  \end{center}
\vspace{-3mm}
  \caption{Comparisons between photometric redshift and spectroscopic redshift for galaxies with spectroscopic redshifts from the literature.  
The insets in the right panels show those for galaxies with $z_{\rm spec} > 1.5$. 
The mean difference between the spectroscopic and photometric redshifts and their 
scatter are also shown. 
}
\label{fig:photz}
\end{figure*}
Extensive spectroscopic surveys have been carried out in the GOODS-North region. 
We cross-matched our $K_{s}$-selected catalog with spectroscopic catalogs from the 
literature. We used the catalogs by \citet{yos10}, \citet{bar08}, \citet{red06}, \citet{tre05}, 
\citet{wir04}, \citet{cow04}, \citet{coh01}, \citet{coh00}, and \citet{daw01}.
As mentioned above, 
the World Coordinate System of  $K_{s}$-band image is based on that of the GOODS 
HST/ACS ver. 2.0 data, which agrees with these spectroscopic catalogs well.  
We first identified a pixel on the $K_{s}$-band image at the source position in the 
spectroscopic catalogs, and then found a $K_{s}$-selected object to which the pixel 
belongs. 
When there were different redshifts from the different catalogs for 
a $K_{s}$-selected object,  we chose one from the most recent catalog. 
Total 2093 $K_{s}$-selected objects are assigned spectroscopic redshifts.

For all objects in the wide and deep catalogs, we also estimated the photometric redshift 
using the $UBVizJHK_{s}$, 3.6$\mu$m, 4.5$\mu$m, and 5.8$\mu$m-band photometry 
described in the previous section. We adopted the standard template-fitting technique 
with population synthesis models. We derived the photometric redshifts with three 
different population synthesis models, namely, GALAXEV \citep{bru03}, PEGASE 
version 2 \citep{fio97}, and \citet{mar05} model, and also used a public photometric 
redshift code, EAZY \citep{bra08} for an independent check of the 
photometric redshift. The results in all four cases are listed in the catalogs.

\begin{figure*}
  \begin{center}
    \FigureFile(140mm,140mm){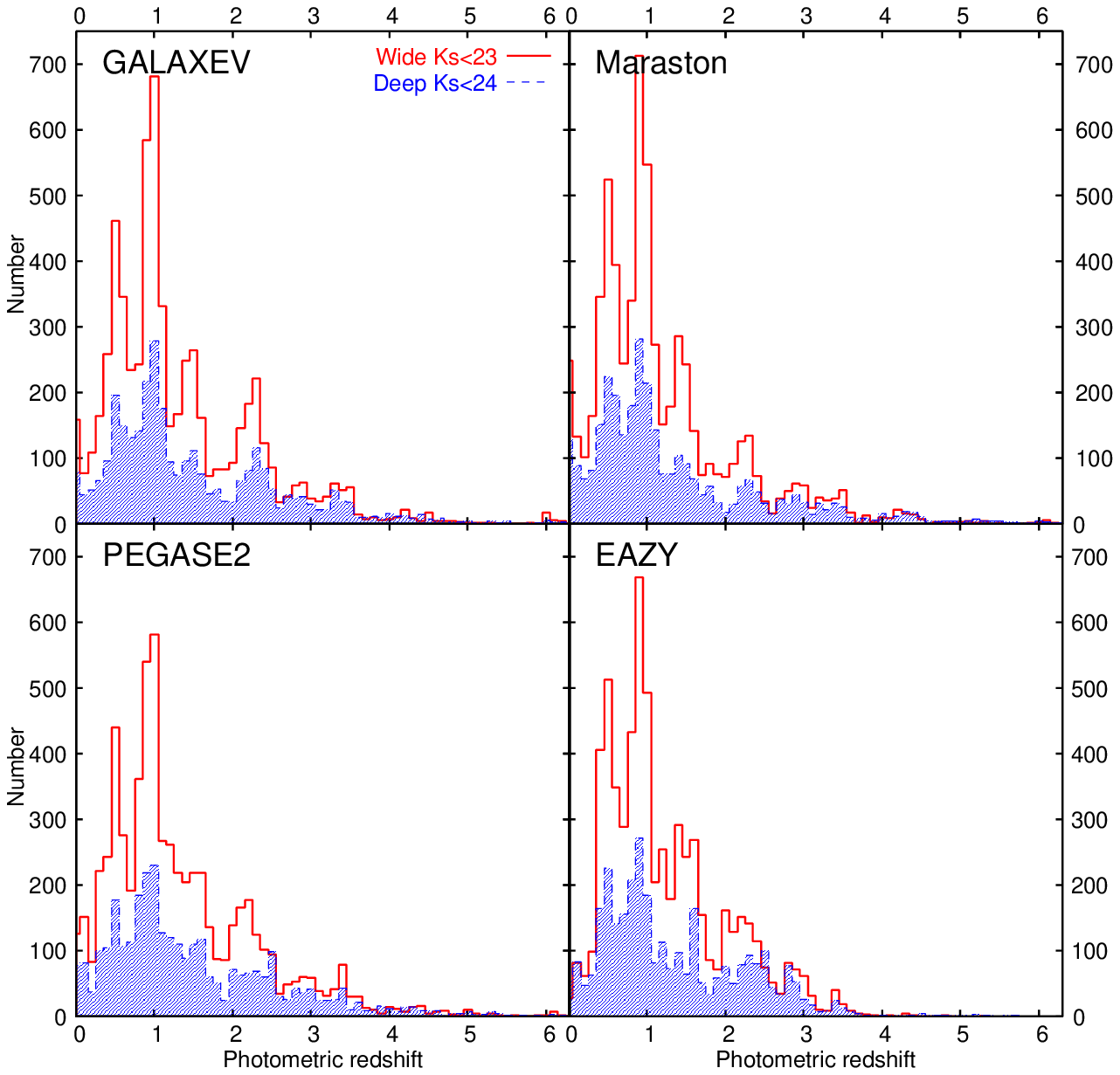}
    %%% \FigureFile(width,height){filename}
  \end{center}
\vspace{-3mm}
  \caption{Photometric redshift distribution for $K_{s}$-selected galaxies in the wide 
(open histogram) and 
deep (shaded histogram) catalogs. 
Four panels show the cases where different population synthesis models 
are used for the photometric redshift. }\label{fig:histzph}
\end{figure*}
The details of the used templates from each population synthesis model are described
in \citet{kaj09}. We here briefly summarize them. 
We adopted exponentially decaying star formation histories and 
assumed Salpeter IMF \citep{sal55} and Calzetti extinction law 
\citep{cal00} for all three models. 
Free parameters of the models are age, 
the star formation timescale $\tau$, dust extinction, 
and metallicity for the GALAXEV and Maraston models. 
The model age is changed from 50 Myr to the age of the universe at the 
observed redshifts, which is calculated  
for a cosmology with H$_{\rm 0}$=70 km s$^{-1}$ Mpc$^{-1}$, 
$\Omega_{\rm m}=0.3$ and $\Omega_{\rm \Lambda}=0.7$.
For the PEGASE2, we used the same template sets as those used in 
\citet{gra06}, where eight exponentially decaying  star formation histories with 
various gas infalling and star formation timescales are assumed. 
The metallicity and dust extinction were calculated in the model 
in a self-consistent way.  We also added to the model templates  the Lyman 
series absorption produced by the intergalactic medium following \citet{mad95}. 
These model SEDs were compared with the observed multi-band photometry and 
the minimum $\chi^{2}$ was found for each object. 

Figure \ref{fig:photz} compares the photometric redshift with 
the spectroscopic redshift for objects with spectroscopic identification for the cases 
with the different population synthesis models. 
The distribution of $\delta z$ ($= z_{\rm photo} - z_{\rm spec}$) and the fraction of 
the catastrophic failure (noted as outliers) 
with $\delta z$/(1+$z_{\rm spec}$) $> 0.5$ are also shown in 
the figure. 
The standard deviation of  $\delta z$/(1+$z_{\rm spec}$) is similar among the 
cases with the different population synthesis models 
($\sim$ 0.08--0.09 for 
all redshift range and $\sim$ 0.11--0.12 for objects at $z_{\rm spec} >1.5$).
Although the photometric redshift accuracy is relatively good in any cases, 
the fraction of the catastrophic failure at $z>1.5$ is slightly larger in the case with 
the Maraston model. 

Figure \ref{fig:histzph} shows the photometric redshift distribution for the cases with 
the different population synthesis models. 
In all cases, the distribution 
shows a primary peak at $z\sim1$ and a secondary peak at $z\sim0.5$  
and a tail to high redshift. 
The peaks in the photometric redshift distribution seem to 
correspond to the large scale structures found by the previous spectroscopic 
surveys at $z=0.48$, 0.51, 0.56, 0.85, and 1.02 
(\cite{coh00}; \cite{wir04}).
Although the distribution is similar among the cases with the different models, 
small systematic differences are also seen. 
In the case with EAZY, for example, 
the fraction of galaxies in the high redshift tail at $z>3$ 
is smaller than 
the other cases, especially at $z\gtrsim 4$. On the other hand, the number of 
galaxies at $z \sim$ 2--2.5 is slightly small in the case with the Maraston model. 
These differences seem to occur in the redshift range where 
 the uncertainty of the photometric redshift is expected to be larger. 
\begin{figure*}
  \begin{center}
    \FigureFile(120mm,120mm){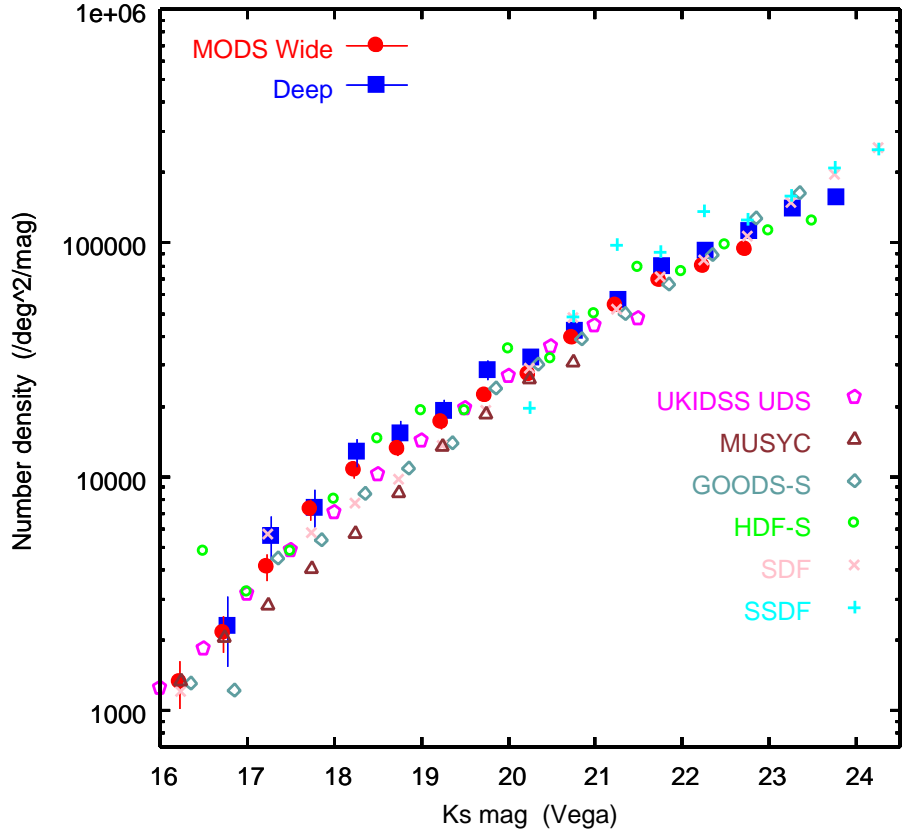}
    %%% \FigureFile(width,height){filename}
  \end{center}
\vspace{-3mm}
  \caption{$K_{s}$-band number counts for objects in the MODS field.
Solid circles and squares show those for the wide and deep fields. 
No correction for the incompleteness or false detection is applied for the MODS data.
Errorbars are based on the Poisson statistics.
Results from other deep NIR surveys are shown for comparison [UKIDSS UDS from 
\citet{wil09}, MUSYC from \citet{qua07}, GOODS-South from \citet{ret10}, 
HDF-South from \citet{lab03}, SDF from \citet{mai01}, and SSDF from \citet{min05}].
}
\label{fig:kcount}
\end{figure*}

\subsection{Spitzer/MIPS 24$\mu$m fluxes}
We measured 24$\mu$m fluxes for the $K_{s}$-selected objects using public 
Spitzer/MIPS 24$\mu$m data of the GOODS survey (DR1+, M. Dickinson et al., 
in preparation). 
As described in \citet{kaj10}, 
since the source confusion occurs for many objects in the MIPS 24$\mu$m data, 
we used the IRAF/DAOPHOT package 
\citep{ste87} for the photometry in order to deal with the blended sources properly, 
following \citet{lef05}.
The DAOPHOT software fitted blended sources in a crowded region simultaneously 
with an PSF, which was empirically constructed from bright isolated point sources in
the MIPS image. 
The source positions in the MOIRCS $K_{s}$-band image are used as a prior for the 
centers of the fitted PSFs in the photometry.
The photometric error was calculated from the background noise, which was estimated 
by measuring sky fluxes at random positions on the MIPS image. 
After the photometry with the DAOPHOT software, 
we also performed the aperture photometry for the PSF-subtracted image outputted by 
the DAOPHOT at the same source positions in order to measure the residuals of the 
PSF fitting. For sources whose fitting residual was larger than the background noise, 
we added the fitting residual to the photometric error. 
In the case where the residual of the fitting is negligible, 
the 5$\sigma$ limiting flux is $\sim$ 20 $\mu$Jy. 

\subsection{Chandra X-ray fluxes}
As described in \citet{yam09}, we also cross-matched our $K_{s}$-selected sample with 
the X-ray point source catalog of the Chandra Deep Field North \citep{ale03}. 
The procedure 
to match the Chandra X-ray source positions in the catalog with the $K_{s}$-selected 
objects was similar with that for the spectroscopic catalogs mentioned above.
We identified a pixel on the $K_{s}$-band image at the X-ray source position and then 
found the $K_{s}$-selected object to which the pixel belongs.
In the case where there is no $K_{s}$-band object at the pixel, we extend the search pixels 
up to three pixel (0.35 arcsec) from the original pixel in order to find the counterpart 
in the $K_{s}$-band image. For most of the Chandra sources,  
the $K_{s}$-band counterparts were actually identified without 
the extended search thanks to the good positional accuracy of the Chandra data.  
Thanks to the depth of the $K_{s}$-band data, 
221 of 226 X-ray sources in the MODS field from the \citet{ale03}'s catalog  
have the $K_{s}$-band counterparts,  
and only 5 sources cannot be identified in the $K_{s}$-band image. 
Of these 5 sources, 4 sources are very faint in X-ray and are detected in only one 
band of the Chandra (full or soft or hard band) at low significance.
Therefore these sources might be spurious sources. 
The other X-ray source is CXOHDFN J123627.5+621218, 
which is detected in all three bands (full-band flux of 3.7 $\times 
10^{-16} erg s^{-1} cm^{-2}$). This source is detected marginally at 3.6 $\mu$m and 
clearly at 4.5 $\mu$m, which suggests that this object has a very red SED. 
For 221 sources with the $K_{s}$-band counterparts, the full, soft, and hard-band 
fluxes are listed in the $K_{s}$-selected catalogs. 

\subsection{Catalog entries}
\label{sec:entry}
The wide and deep $K_{s}$-selected catalogs are publicly available on the MODS 
web site\footnote{http://www.astr.tohoku.ac.jp/MODS/}. 
It should be noted that all magnitudes in the catalogs are in AB 
magnitude system.
The conversion from the Vega system to the AB system in the MOIRCS $JHK_{s}$ bands 
are $J_{\rm AB} = J_{\rm Vega} + 0.915$, $H_{\rm AB} = H_{\rm Vega} + 1.354$, and 
$K_{s {\rm AB}} = K_{s {\rm Vega}} + 1.834$, respectively.
The catalog entries are as follows.
\begin{table*}
  \caption{$K_{s}$-band number counts for the wide and deep fields.
Errors of the number density are based on the Poisson statistics.}
\label{tab:first}
  \begin{center}
    \begin{tabular}{lrrr|rrr}  
   \hline
   & \multicolumn{3}{c}{Wide (103.3 arcmin$^{2}$)} \vline& \multicolumn{3}{c}{Deep (28.2 arcmin$^{2}$)}\\
   $K_{s}$ mag & Number & Number density & error & Number & Number density & error \\
   (Vega) & & (deg$^{-2}$ mag$^{-1}$) &  (deg$^{-2}$ mag$^{-1}$) & & (deg$^{-2}$ mag$^{-1}$) & (deg$^{-2}$ mag$^{-1}$) \\ 
\hline
  16.25 & 19 & 1223.9 & 303.7 & 1 & 255.0 & 255.0 \\
  16.75 & 31 & 2160.0 & 387.9 & 9 & 2295.0 & 765.0 \\
  17.25 & 59 & 4111.0 & 535.2 & 22 & 5610.1 & 1196.1 \\
  17.75 & 104 & 7246.5 & 710.6 & 29 & 7395.1 & 1373.2 \\
  18.25 & 154 & 10730.4 & 864.7 & 50 & 12750.1 & 1803.1 \\
  18.75 & 190 & 13238.8 & 960.4 & 60 & 15300.2 & 1975.2 \\
  19.25 & 245 & 17071.0 & 1090.6 & 75 & 19125.2 & 2208.4 \\
  19.75 & 320 & 22296.8 & 1246.4 & 112 & 28560.3 & 2698.7 \\
  20.25 & 391 & 27244.0 & 1377.8 & 127 & 32385.3 & 2873.7 \\
  20.75 & 565 & 39367.9 & 1656.2 & 164 & 41820.4 & 3265.6 \\
  21.25 & 777 & 54139.5 & 1942.2 & 225 & 57375.6 & 3825.0 \\
  21.75 & 999 & 69608.0 & 2202.3 & 311 & 79305.8 & 4497.0 \\
  22.25 & 1137 & 79223.5 & 2349.5 & 362 & 92311.0 & 4851.8 \\
  22.75 & 1346 & 93786.1 & 2556.3 & 438 & 111691.2 & 5336.8 \\
  23.25 & & & & 550 & 140251.5 & 5980.3 \\
  23.75 & & & & 614 & 156571.6 & 6318.7 \\
\hline
\label{tab:count}
    \end{tabular}
  \end{center}
\end{table*}

\begin{itemize}
\item ID --- sequential identification number in order as outputted by SExtractor. IDs 
in the wide and deep catalogs are independently assigned (see ID\_[deep/wide] below). 

\item X\_PIXEL, Y\_PIXEL --- pixel coordinates in the mosaiced $K_{s}$-band image.

\item RA, Dec --- right ascension and declination coordinates (J2000), 
which are based on the GOODS HST/ACS ver. 2.0 data.

\item K\_AUTO --- $K_{s}$-band Kron magnitude (MAG\_AUTO from SExtractor) in AB 
magnitude. 

\item K\_AUTO\_COR --- $K_{s}$-band Kron magnitude corrected for fluxes missed from 
the Kron aperture (assuming the surface brightness profile of point sources, see Section \ref{sec:detect} and Appendix \ref{sec:apcor} for details).

\item SNR\_AUTO --- signal to noise ratio for the $K_{s}$-band Kron magnitude. 

\item rad\_AUTO --- aperture radius used for the $K_{s}$-band Kron magnitude 
(2.5 $\times$ Kron radius) in pixel unit. 
The circularized value is listed.

\item rad\_half --- $K_{s}$-band circularized half light radius in pixel unit.

\item flag --- extraction flags from SExtractor

\item \{UBVizJHKs, 3.6, 4.5, 5.8, 8.0\}\_ap --- aperture magnitude in AB magnitude in 
each band. The aperture with 1.2 (1.0) arcsec diameter is basically 
used for the wide (deep) catalog. 

\item  \{UBVizJHKs, 3.6, 4.5, 5.8, 8.0\}\_err --- photometric error for the aperture magnitude 
in each band. Negative error indicates that the aperture magnitude is the 2$\sigma$ 
upper limit.

\item f\_24um, err\_24um --- MIPS 24$\mu$m flux in $\mu$Jy and associated error.

\item flux\_X, flux\_XSB, flux\_XHB --- Chandra X-ray fluxes in full, soft, and hard bands, 
respectively, in erg s$^{-1}$ cm$^{-2}$. 

\item z\_spec --- spectroscopic redshift from the literature.

\item refer\_spec --- reference for the spectroscopic redshift. 1: \citet{yos10}, 2: 
\citet{bar08}, 3: \citet{red06}, 5: \citet{tre05}, 6: \citet{wir04}, 7: \citet{cow04}, 
8: \citet{coh01}, 9: \citet{coh00}, 10: \citet{daw01}.

\item IDref\_spec --- ID in the spectroscopic catalog. 

\item zph\_BC, zph\_MR, zph\_PG --- photometric redshifts with the population 
synthesis models of GALAXEV, Maraston, and PEGASE2. 

\item zph\_EZ --- photometric redshift outputted by EAZY.

\item wht\_\{UBViz\} --- weight value (inverse variance per pixel) 
at the object position in each band. 

\item exp\_\{JHKs, 3.6, 4.5, 5.8, 8.0\} --- exposure time at the object position 
in seconds in each band.

\item exp\_24um --- exposure time at the object position 
in seconds in the MIPS 24$\mu$m image.

\item exp\_\{X, XSB, XHB\} --- effective exposure time at the object position 
in seconds in the Chandra full, soft, and hard-band images, respectively.

\item ID\_deep, ID\_wide --- ID of the corresponding object in the other 
(deep or wide) catalog. 

\end{itemize}

\section{Analysis}
We here present the $K_{s}$-band number counts and distribution 
of NIR colors, and demonstrate some selection techniques with the NIR colors 
for high redshift galaxies. 
In appendix \ref{sec:comphot}, we also compare the $K_{s}$-band total magnitude 
and colors between the K$_{s}$ band and the other $U$-to-8.0$\mu$m bands in 
the MODS catalogs with those in other studies to check our photometry. 

\subsection{$K_{s}$-band Number Counts}
Using the $K_{s}$-selected catalogs described above, 
we constructed $K_{s}$-band number counts.
 Figure \ref{fig:kcount} and Table \ref{tab:count} 
show the number counts for objects in the wide and deep fields.   
We limited the samples to $K_{s}<23$ ($K_{s}<24$) for objects in the wide (deep) 
field, where the detection completeness is very high ($\gtrsim 90$\% at least for 
 point sources) and the spurious detection is negligible as seen in 
Figure \ref{fig:comp}. 
No correction for the incompleteness or the false detection is applied.
The number counts are calculated in 0.5 mag bins and error bars are based on 
Poisson statistics.
For comparison, those of other deep NIR surveys, namely, the UKIDSS UDS 
with UKIRT/WFCAM \citep{wil09}, the MUSYC deep NIR fields with CTIO 4m/ISPI
\citep{qua07},  the GOODS-South with VLT/ISAAC \citep{ret10}, 
the FIRES HDF-South with VLT/ISAAC \citep{lab03}, Subaru Deep Field (SDF) with 
Subaru/CISCO \citep{mai01}, and Subaru Super Deep Field (SSDF) with 
Subaru/IRCS+AO \citep{min05}, are also shown in the figure.
$K_{s}$-band counts of the MODS agree well with the results of other surveys 
especially at faint magnitudes. 
The MODS number counts are slightly higher than those of other surveys at 
$K_{s}\sim $ 18--19 and 
it is also seen that 
the number counts in the MODS deep field is sightly higher than those in the wide 
field. 
These might be due to the large scale structures around the HDF-N 
(\cite{coh00}; \cite{wir04}) which the GT-2 field includes as mentioned above.

The logarithmic slope $d(\log{N})/dm$ of the MODS number counts seems to 
decrease with magnitude. 
For example, the best-fit slope to the number counts at $20<K_{s}<24$ 
is $\alpha = 0.20 \pm 0.04$, while the slope becomes steeper at $K_{s}\lesssim 18$.
The change in the logarithmic slope of the $K$-band counts at $K \sim $ 17--18 
has been reported by previous studies (e.g., \cite{cri03}; \cite{cri09}; \cite{qua07}). 
The slope at the faint magnitude of $K_{s}>20$ is consistent with the results of 
the FIRES survey (\cite{lab03}; \cite{for06}), although it is slightly flatter than 
those of the GOODS-S and SDF (\cite{ret10}; \cite{mai01}). 
If we divide the sample at $K_{s}\sim 21$ and estimate the slope of the number 
counts separately, 
the slope is $\alpha = 0.22 \pm 0.06$ at $18<K_{s}<21$ and $\alpha = 0.18 \pm 0.05$ 
at $21<K_{s}<24$. 
Although the uncertainty is rather large, 
the flattening of the slope with magnitude may continue even at  $K\gtrsim 19$.  
\begin{figure}
  \begin{center}
    \FigureFile(80mm,90mm){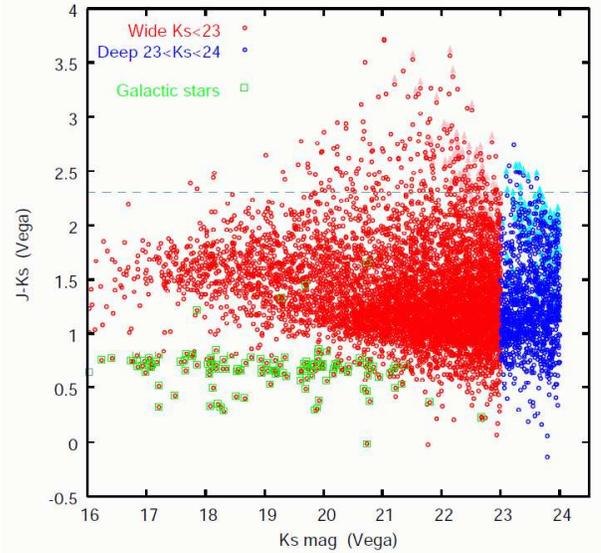}
    %%% \FigureFile(width,height){filename}
  \end{center}
\vspace{-3mm}
  \caption{$J-K_{s}$ vs. $K_{s}$ color-magnitude diagram for objects in the MODS field.
Red circles show objects with $K_{s}<23$ in the wide field, while blue ones show 
objects with $23<K_{s}<24$ in the deep field.
Objects with S/N $<$ 2 in $J$ band are plotted 
at their 2$\sigma$ lower limit of the $J-K_{s}$ color with arrows.
Open squares represent spectroscopically confirmed Galactic stars.
Horizontal dashed line shows $J-K_{s}=2.3$, which is used as for the DRG selection.
}\label{fig:jk}
\end{figure}

\subsection{NIR Color distribution}
Figure \ref{fig:jk} shows $J-K_{s}$  vs.  $K_{s}$  color-magnitude diagram for objects with 
$K_{s}<23$ in the wide field and those with $23<K_{s}<24$ in the deep field. 
Spectroscopically confirmed stars are shown as squares in the figure. 
These stars form a upper envelope at $J-K_{s} \sim 0.9$ in their color-magnitude distribution, 
which is consistent with the expected locus of Galactic stars. 
Objects with S/N $<$ 2 in $J$ band are plotted at their 2$\sigma$ lower 
limit of the $J-K_{s}$ color as symbols with arrows. 
Thanks to our deep $J$-band data (the 2$\sigma$ limit of $J \sim 25.2$ for the wide 
field and $J \sim 26.1$ for the deep field), we can do the color measurement up to 
 rather red $J-K_{s}$ colors.
\begin{figure*}
  \begin{center}
    \FigureFile(95mm,95mm){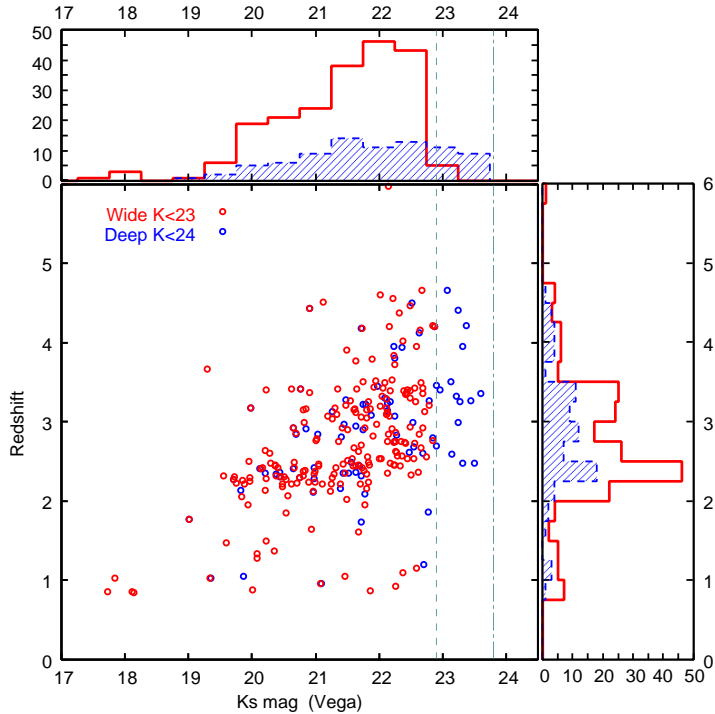}
    %%% \FigureFile(width,height){filename}
  \end{center}
\vspace{-3mm}
  \caption{Redshift vs. $K_{s}$-band magnitude for DRGs in the MODS field.
Red symbols show DRGs with $K_{s}<23$ in the wide field, while blue symbols 
represent DRGs with $K_{s}<24$ in the deep field.
Upper panel shows the $K_{s}$-band number counts of the DRGs, and 
right panel shows the redshift distribution of these objects.
Vertical dashed and dotted-dashed lines 
represent the $K_{s}$-band limit for the 
DRGs in the wide and deep fields, respectively.
}\label{fig:drg}
\end{figure*}
\begin{figure}
  \begin{center}
    \FigureFile(80mm,150mm){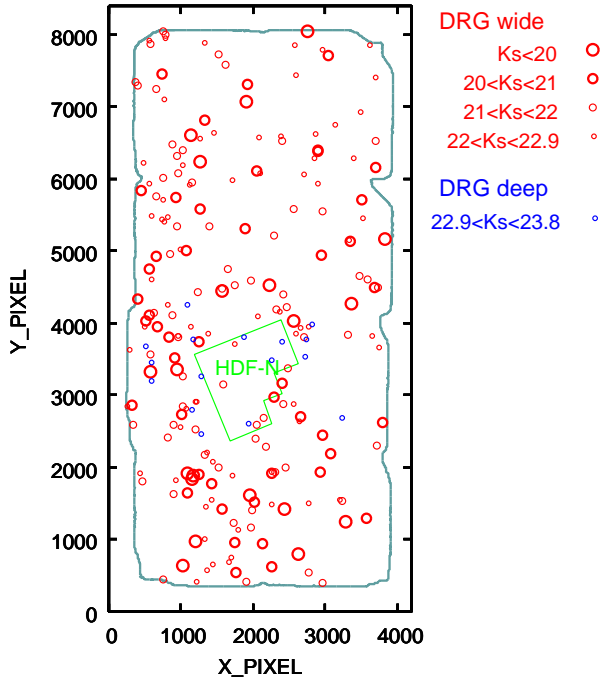}
    %%% \FigureFile(width,height){filename}
  \end{center}
\vspace{-3mm}
  \caption{Spatial distribution of DRGs in the MODS field. 
The size of the symbols is scaled 
according to apparent magnitudes in the $K_{s}$-band. 
Red symbols show the DRGs with $K_{s}<22.9$ in the wide field, while 
blue symbols represent those with $22.9<K_{s}<23.8$ in the deep field.
Thick solid line shows the survey area for which our catalog was constructed 
(t$_{\rm exp} > 3600$ sec in the $H$-band).   
The original HDF-N field is also shown as thin line.  
}\label{fig:xydrg}
\end{figure}
 For example, we can measure 
 objects with $J-K_{s} \sim 3$ down to $K_{s}\sim22.2$ ($K_{s}\sim23.1$) for the wide (deep) field.  
The dashed line in the figure represents $J-K_{s} = 2.3$, which is used as the limit for 
the Distant Red Galaxies (DRGs, \cite{fra03}). 
We can select DRGs with $J-K_{s}>2.3$ down to $K_{s}\sim 22.9$ for the wide field and 
$K_{s}\sim 23.8$ for the deep field, and detected 208 DRGs in the wide field and 
81 DRGs in the deep field (65 are overlapping). 
Figure \ref{fig:drg} shows the distribution of the $K_{s}$-band magnitude and redshift for 
the DRGs. In the figure, we used the spectroscopic redshifts (18 galaxies) if available, 
otherwise the photometric redshifts with the GALAXEV model. 
The $K_{s}$-band number counts of the DRGs seem to turnover at $K_{s}\sim22$ 
and do not increase with magnitude at $K_{s} \gtrsim 22$ 
as found in \citet{kaj06}. 
The DRGs lie mainly at $2\lesssim z \lesssim 3.5$, and there are tails in the 
redshift distribution at $z\sim$ 0.5--2 and $z\sim$ 3.5--4.5, which is consistent with 
the results of other deep surveys (e.g., \cite{for06}; \cite{qua07}; \cite{qua08}).
At low redshift ($z\lesssim 2$), the number of DRGs does not increase with magnitude, 
and that of faint DRGs is relatively small.
A similar trend is also seen at higher redshift in the magnitude range we investigated.
On the other hand, the brightest $K_{s}$-band magnitude of the DRGs at each redshift 
gradually shifts to fainter values with redshift.  
As a result, 
the redshift distribution of the DRGs shifts to higher redshift with $K_{s}$-band magnitude. 
The DRGs at relatively low redshift dominate at bright magnitude ($K_{s} \lesssim 19$), 
which is also consistent with the result of \citet{con07}.

Figure \ref{fig:xydrg} shows the spatial distribution of the DRGs in the MODS field.
As found by \citet{ich07} for those in the GT-2 field, 
these DRGs seem to be strongly clustered. 
For example, 
several bright DRGs concentrate at (X$\sim$1200, Y$\sim$1900).  
One of these galaxies 
has $z_{\rm spec}=3.661$, while five of the other six DRGs show 
$z_{\rm phot} = $ 2.2--2.5 (the other one has $z_{\rm phot} \sim$ 1.3). 
On the other hand, 
 there are few bright DRGs in the original HDF-N field as shown in \citet{kaj06}.
At $\sim$ 
2.5 arcmin scale ($\sim$ 1300 pixels), 
there are other similar voids of bright DRGs in the MODS field.
The field-to-field variance of the number density of these galaxies 
seems to be large at the scale. 

\begin{figure*}
  \begin{center}
    \FigureFile(150mm,100mm){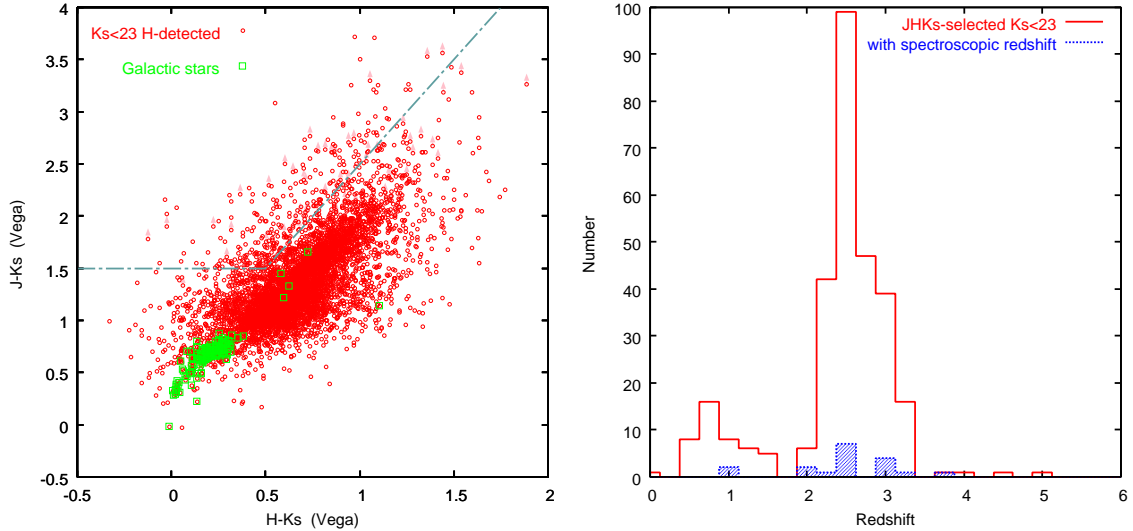}
    %%% \FigureFile(width,height){filename}
  \end{center}
\vspace{-3mm}
  \caption{{\bf left:} 
$J-K_{s}$ vs. $H-K_{s}$ two-color diagram for objects in the MODS field.
Only objects with $K_{s}<23$ and S/N $>$ 2 in $H$ band located 
in the wide field which are detected in $H$ band are plotted. 
Objects with S/N $<$ 2 in $J$ band are plotted 
at their 2$\sigma$ lower limit of the $J-K_{s}$ color with arrows.
Open squares show spectroscopically confirmed stars.
Dashed-dotted line represents the selection criterion for Balmer break galaxies 
at $z\sim2.5$ introduced by \citet{kaj06b}, namely,  
$J-K_{s} > 2(H-K_{s})+0.5$ and $J-K_{s}>1.5$.
{\bf right:} 
 Redshift distribution of the Balmer break galaxies selected by the $JHK_{s}$ color 
criterion. Open histogram shows all selected objects with $K_{s}<23$, and shaded 
histogram shows those with spectroscopic identification.
}
\label{fig:jhk}
\end{figure*}
The left panel of 
Figure \ref{fig:jhk} shows $J-K_{s}$ vs. $H-K_{s}$ color-color diagram for objects with $K_{s}<23$ 
in the MODS field. 
We show only objects with S/N $>$ 2 in $H$ band in the figure.
Again, squares represent the spectroscopically confirmed stars. 
They lie mainly at $0.3 \lesssim J-K_{s} \lesssim 0.9$ and $0 \lesssim H-K_{s} \lesssim 0.3$, 
and there are also some red stars.
Their distribution in the color-color diagram is consistent with that expected for 
Galactic stars, which indicates that 
the zero points of our $J$, $H$, and $K_{s}$-band data have been
 correctly determined.
The distribution of galaxies in the $J-K_{s}$ vs. $H-K_{s}$ diagram is also consistent with 
those seen in other NIR surveys (e.g., \cite{for06}; \cite{ret10}). 
For an example of a color selection technique for high-redshift galaxies in the 
color-color diagram, 
we show the selection criterion for Balmer break galaxies at $z\sim2.5$ introduced by 
\citet{kaj06b} in the left panel of Figure \ref{fig:jhk} (dotted-dashed line).
This criterion is $J-K_{s} > 2(H-K_{s})+0.5$ and $J-K_{s}>1.5$, and is met while the Balmer/4000  
\AA\ break of galaxies falls between $J$ and $H$ bands (i.e., $2\lesssim z \lesssim3$), 
whereas most of the foreground and background galaxies would not meet such 
a criterion. The right panel of Figure \ref{fig:jhk} shows the redshift distribution of 
objects selected by this criterion. 
Again we used the spectroscopic redshifts if available, 
otherwise the photometric ones with the GALAXEV model. 
Most of these galaxies  lie at $2\lesssim z \lesssim 3.2$ showing a strong peak at 
$z\sim2.5$, although there is a relatively small contamination ($\sim 15$ \%) 
from low-redshift galaxies.  Thus, the deep $JHK_{s}$-band data of the MODS allow
us to sample the SEDs around the Balmer/4000 \AA\  break 
 of galaxies at $2\lesssim z \lesssim 3$ with high accuracy, which is important for 
investigating the stellar population of these high-redshift galaxies.

\section{Summary}
In this paper, we presented the deep $JHK_{s}$-band imaging data of 
the MODS obtained with MOIRCS on the Subaru telescope. 
The data cover an area of 103.3 arcmin$^{2}$ in the GOODS-North region and  
reach the 5$\sigma$ total limiting magnitudes for point sources 
of $J=23.9$, $H=22.8$, and $K_{s}=22.8$.  
In 28.2 arcmin$^{2}$ of the survey area, 
the data reach the 5$\sigma$ depths of $J=24.8$, $H=23.4$, $K_{s}=23.8$. 
The World Coordinate System of the reduced images is based on the public 
HST/ACS ver. 2.0 data of the GOODS.
The image quality of the combined images is characterized by 
a  PSF with FWHM of $\sim $ 0.53--0.60 arcsec for the wide field and $\sim$ 
0.45--0.49 arcsec for the deep field, respectively.
For the color measurements, 
we also provided the PSF-matched mosaic images whose FWHM of the PSFs 
are 0.6 arcsec for the wide field and 0.5 arcsec for the deep field.
The $K_{s}$-band detection completeness for the point sources is $\sim$ 90 \% 
at $K_{s}\sim23$ for the wide field and at $K_{s}\sim24$ for the deep field. 
The spurious sources are also negligible down to these completeness limits, although
the false detection rate increases rapidly below the limits. 

Combining the multi-wavelength public data taken with the HST, Spitzer, and other 
ground-based telescopes in the GOODS field with the MODS data,
we constructed the multi-wavelength photometric catalogs of $K_{s}$-selected 
sources. 
Total 9875 and 3787 objects are listed in the wide and deep catalogs. 
The catalogs also include the Spitzer/MIPS 24 $\mu$m fluxes, Chandra 
X-ray fluxes, and redshifts of the $K_{s}$-selected objects.
The comparisons between the spectroscopic and photometric redshifts suggest  
that the photometric redshift accuracy is  $\delta z/(1+z_{\rm spec}) \lesssim 0.1$
with $\sim$ 3\% outliers for galaxies with spectroscopic redshifts 
($\sim $ 4--12\% outliers for  objects with $z_{\rm spec} > 1.5$).

Using the catalogs, we examined the $K_{s}$-band number counts and NIR color 
distribution. The $K_{s}$-band number counts in the MODS field are consistent with 
those in other general fields and show the logarithmic slope of $d(\log{N})/dm \sim 0.2$  
 at $K\gtrsim 20$. 
The NIR color distribution of spectroscopically confirmed stars indicates 
that the zero points of the MODS data have been correctly determined.
We also demonstrated some selection techniques for high-redshift galaxies 
with the NIR colors. 

The MODS data are one of the deepest NIR 
imaging data  over a area of $\sim$ 30 to $\sim$ 100 arcmin$^{2}$ to date, 
especially in the $K_{s}$ band.
These data sample an important part of the SEDs of galaxies at $1\lesssim z \lesssim 4$ 
to study their stellar population.
The high image quality of the data is very useful for the deconvolution of 
 other imaging data with poorer spatial resolution, while their depth in the NIR 
is essential for the identification of objects detected in the multi-wavelength data,  
 which have been obtained by the Chandra, Spitzer, Herschel, VLA, and so on. 
These imaging data and  $K_{s}$-selected source catalogs are publicly available 
on the MODS 
web site (http://www.astr.tohoku.ac.jp/MODS/).

\bigskip

We thank an anonymous referee for very helpful suggestions and comments. 
We also thank Kevin Bundy for kindly providing their KGOODS-N v1.0 catalog.  
This study is based on data collected at Subaru Telescope, which is operated by
the National Astronomical Observatory of Japan. 
This work is based in part on observations made with the Spitzer Space
Telescope, which is operated by the Jet Propulsion Laboratory, 
California Institute of Technology under a contract with NASA.
Some of the data presented in this paper were obtained from the Multi-mission
Archive at the Space Telescope Science Institute (MAST).
STScI is operated by the Association of Universities for Research in
Astronomy, Inc., under NASA contract NAS5-26555.
Support for MAST for non-HST data is provided by the NASA Office of
Space Science via grant NAG5-7584 and by other grants and contracts.
Data  reduction and analysis 
were carried out on common use data analysis computer system 
 at the Astronomy Data Center, ADC, of the National Astronomical 
Observatory of Japan.
IRAF is distributed by the National Optical Astronomy Observatories,
which are operated by the Association of Universities for Research
in Astronomy, Inc., under cooperative agreement with the National
Science Foundation.

%Acknowledgement should be placed at end of main text.
%(NOT after the Appendix.)

\appendix
\section{Simulations for estimating the aperture corrections}
\label{sec:apcor}
\begin{figure}
  \begin{center}
    \FigureFile(75mm,160mm){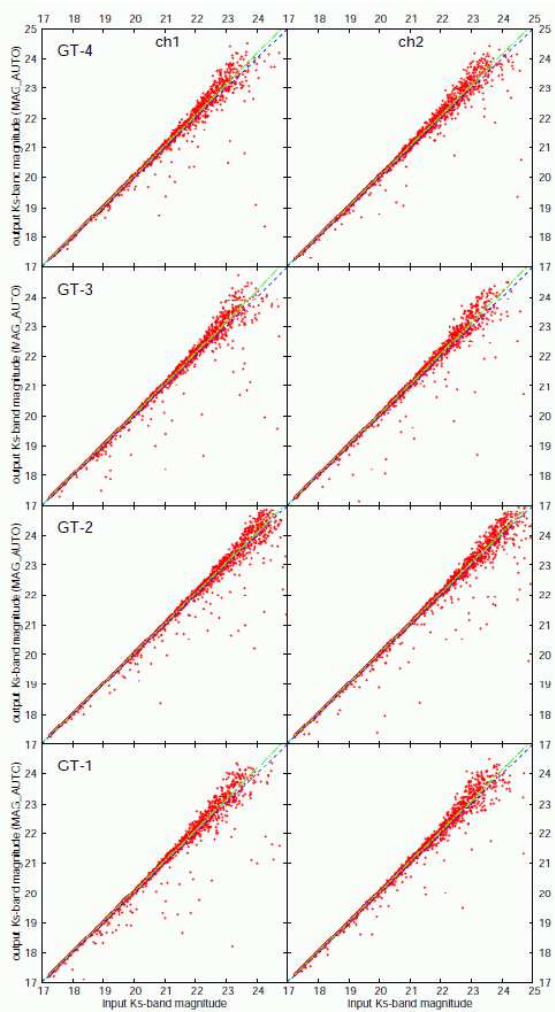}
    %%% \FigureFile(width,height){filename}
  \end{center}
\vspace{-3mm}
  \caption{Comparison between the input $K_{s}$-band total magnitude and the 
measured MAG\_AUTO for artificial point sources in the simulations  
for each chip and field. Dotted-dashed line shows the best-fit second-order 
polynomial function. 
}
\label{fig:simauto}
\end{figure}

In this section, we investigate how the MAG\_AUTO from SExtractor misses 
fluxes as a function of magnitude, using simulations with the IRAF/ARTDATA package. 
Although such aperture correction could depend on not only magnitude but also 
size, surface brightness profile, and morphology of objects, we here examine 
the aperture correction for point sources as a guide. 
The aperture correction for point sources can be considered to be the 
minimum correction. 
The procedures are the same as in the measurement of the detection completeness 
in Section \ref{sec:detect}. 
An artificial point source with the corresponding PSF was added at random
position in the $K_{s}$-band image of each chip and field. 
We then ran the SExtractor with the same detection parameters 
and compared the input magnitude and the output MAG\_AUTO. 
The results are shown in Figure \ref{fig:simauto}. 
The MAG\_AUTO is fainter by $\sim$ 0.02--0.03 mag 
than the input total magnitude at $K\sim$ 17--18 and 
becomes fainter by $\sim$ 0.15 mag near the 5$\sigma$ limiting magnitude 
($K\sim 23$ for the GT-1, 3, and 4 fields, and $K\sim 24$ for the GT-2 fields).
The aperture correction seems to be slightly smaller for the GT-2 field data, which 
is $\sim$  1 mag deeper and has higher image quality. 
We fitted the difference between the input total magnitude and the output MAG\_AUTO 
(i.e., the aperture correction) as a function of the {\it measured} MAG\_AUTO with 
a second-order polynomial function. 
We used the results to correct the MAG\_AUTO and listed the corrected MAG\_AUTO 
as MAG\_AUTO\_COR in our catalogs (Section \ref{sec:entry}).  
\begin{figure}
  \begin{center}
    \FigureFile(75mm,160mm){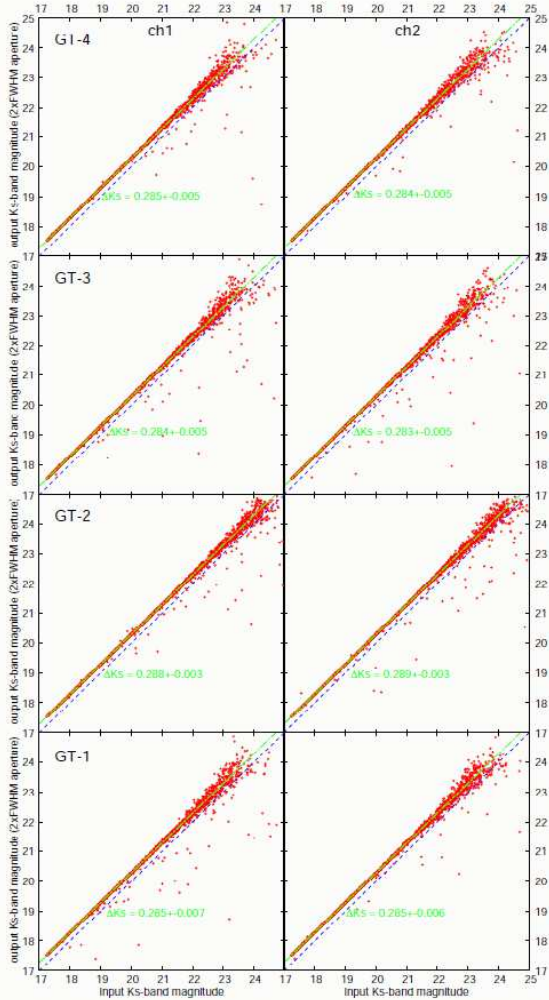}
    %%% \FigureFile(width,height){filename}
  \end{center}
\vspace{-3mm}
  \caption{Comparison between the input $K_{s}$-band total magnitude and the 
aperture magnitude measured with a aperture diameter of 2 $\times$ FWHM of the PSF 
in the simulations for each chip and field. 
Dotted-dashed line represents the result of the linear fitting for 
the data at $K_{s} < 20$. 
}
\label{fig:simap}
\end{figure}

In the simulations, we also measured the aperture magnitude for the artificial 
point sources with a aperture diameter of 2 $\times$ FWHM of the PSF to 
estimate the aperture correction for the limiting magnitude measured with the 
aperture in Section \ref{sec:detect}.
Figure \ref{fig:simap} compares the aperture magnitude measured with 
the 2 $\times$ FWHM diameter with the input total magnitude. 
The difference between these two magnitudes is nearly constant over a wide 
range of the magnitude, especially at bright magnitude. 
We fitted the difference at the input $K_{s}<20$, 
assuming that it is constant over magnitude. 
 The resulting aperture correction is similar among the different chips and fields and 
is $\sim$ 0.28--0.29 mag. 
The total limiting magnitudes for point sources are expected to be brighter by these 
values than those estimated from the background fluctuation which was measured 
with a aperture diameter of 2 $\times$ FWHM. 
The corrected limiting magnitudes for point sources are also listed in Table \ref{tab:field}.

\section{Comparison of the photometry with other studies in the GOODS-North}
\label{sec:comphot}
In order to check the photometric accuracy, 
we here compare the $K_{s}$-band total magnitude 
and colors between the $K_{s}$ band and the other $U$-to-8.0$\mu$m bands 
in our catalogs with other studies in the GOODS-North field.
The procedure to match sources in the catalogs of other studies with 
those in the MODS catalog is the same as that for the spectroscopic catalogs 
(Section \ref{sec:redshift}). 
We excluded objects with the deblended flag (from SExtractor) in our catalog from 
the comparisons in order to avoid the effects of the different spatial resolution of 
the images used in the different studies on the source detection.  
\begin{figure*}
  \begin{center}
    \FigureFile(130mm,90mm){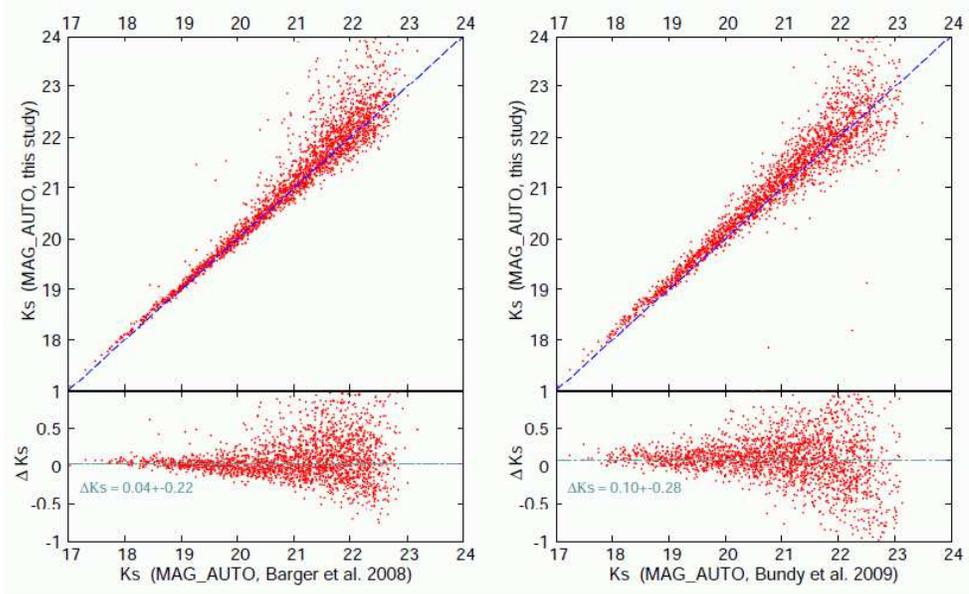}
    %%% \FigureFile(width,height){filename}
  \end{center}
\vspace{-3mm}
  \caption{Comparisons of the total $K_{s}$-band magnitudes (MAG\_AUTO) in this 
study with those by \citet{bar08} (left) and by \citet{bun09} (right). 
}
\label{fig:kauto}
\end{figure*}

\begin{figure*}
  \begin{center}
    \FigureFile(130mm,130mm){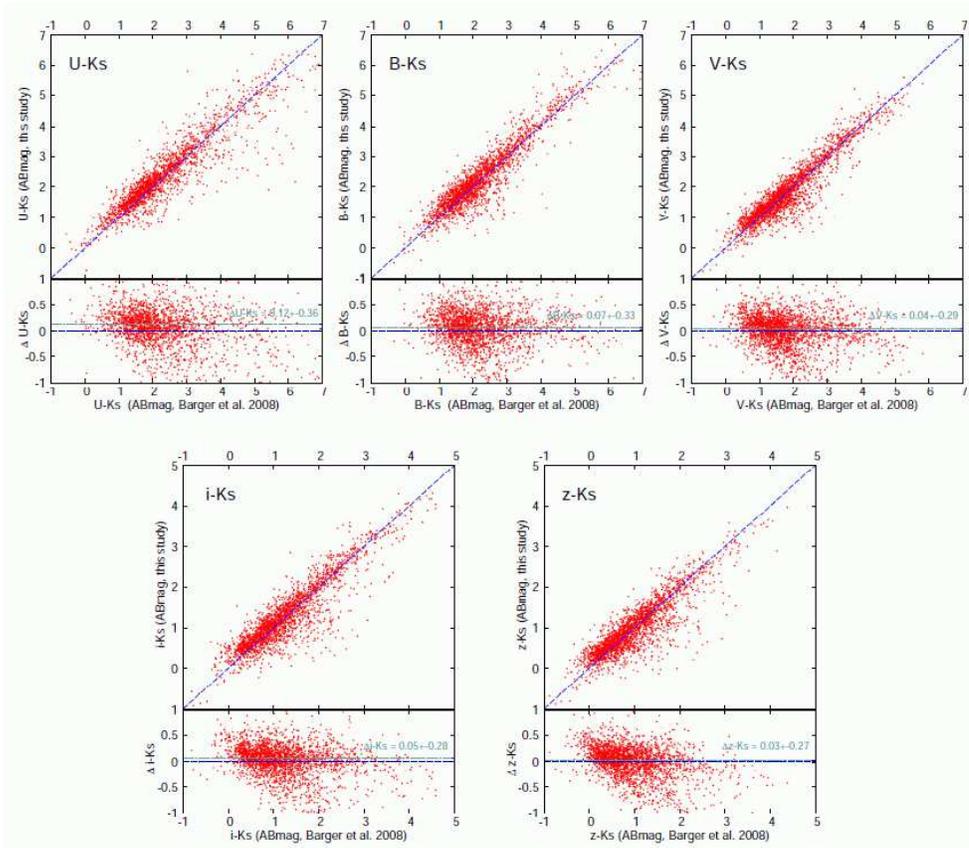}
    %%% \FigureFile(width,height){filename}
  \end{center}
\vspace{-3mm}
  \caption{Comparisons of $U-K_{s}$, $B-K_{s}$, $V-K_{s}$, $i-K_{s}$, and $z-K_{s}$ 
colors between this study and \citet{bar08}. It should be noted that 
the colors from \citet{bar08} are measured with the Kron aperture (MAG\_AUTO), 
while those for the MODS are measured with a 1.2 arcsec diameter aperture. 
}
\label{fig:color1}
\end{figure*}

Figure \ref{fig:kauto} compares the $K_{s}$-band total magnitudes  
in the MODS catalog with those in \citet{bar08} and \citet{bun09}. 
We here used the original (not aperture-corrected) MAG\_AUTO for 
direct comparisons with these studies. 
\citet{bar08} obtained deep $K_{s}$-band data with CFHT/WIRCam and 
calibrated the WIRCam data using their MOIRCS data \citep{wan09}. 
On the other hand, 
\citet{bun09} obtained relatively shallow $K_{s}$-band data with Subaru/MOIRCS and 
carried out the photometric calibration of their MOIRCS data using the 
Palomar/WIRC data \citep{bun05}. 
Both studies used the MAG\_AUTO from SExtractor as the total magnitude.
In the figure, the $K_{s}$-band total magnitudes in the MODS catalog 
are slightly fainter on average than \citet{bar08} (by 0.04 mag) and \citet{bun09} 
(by 0.10 mag), while the overall agreements are relatively good in the both cases.
Considering the zero-point uncertainty, 
the total magnitudes are consistent among the studies.

\begin{figure}
  \begin{center}
    \FigureFile(80mm,110mm){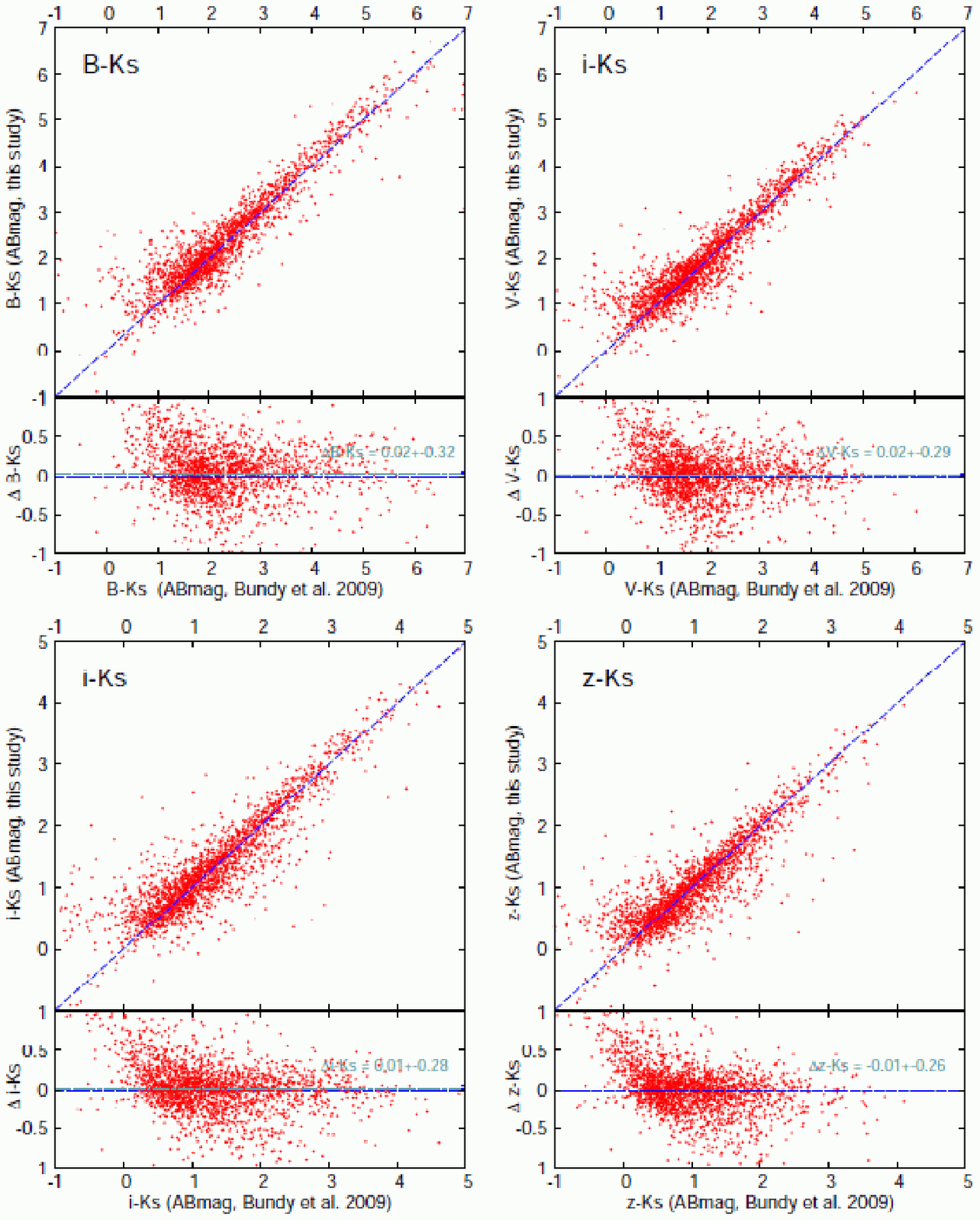}
    %%% \FigureFile(width,height){filename}
  \end{center}
\vspace{-3mm}
  \caption{Comparisons of $B-K_{s}$, $V-K_{s}$, $i-K_{s}$, and $z-K_{s}$ 
colors between this study and \citet{bun09}. While the colors in this study 
are measured with a 1.2 arcsec diameter aperture, those from \citet{bun09} are 
measured with a 2.0 arcsec diameter aperture. 
}
\label{fig:color2}
\end{figure}

Figure \ref{fig:color1} shows comparisons of $U-K_{s}$, $B-K_{s}$, $V-K_{s}$, 
$i-K_{s}$, and $z-K_{s}$ colors between \citet{bar08} and this study. 
It should be noted that \citet{bar08} used MAG\_AUTO for the color measurements, 
while the aperture diameter of 1.2 arcsec is used for the MODS data 
(the wide catalog). Despite the difference of the aperture size, these colors 
agree well, especially in the colors between the MOIRCS $K_{s}$ band and 
the HST/ACS $BViz$ bands.  
For $U$-band photometry, we used the aperture correction with the surface 
brightness profiles of the  $B$-band image, and the photometric error tends to 
be relatively large. 
The average difference of 0.12 mag in the $U-K_{s}$ color  between \citet{bar08} and 
this study is expected from the uncertainty, and it may be due to 
some small systematic effect in the aperture correction. 
On the other hand, Figure \ref{fig:color2} compares the $B-K_{s}$, $V-K_{s}$, $i-K_{s}$, 
and $z-K_{s}$ colors between \cite{bun09} and our catalog. 
We again used the aperture magnitude with a 1.2 arcsec diameter for the MODS data 
and adopted the aperture magnitude with a 2.0 arcsec diameter from the catalog of 
\citet{bun09}. An excellent agreement can be seen in all the colors.

\begin{figure}
  \begin{center}
    \FigureFile(80mm,110mm){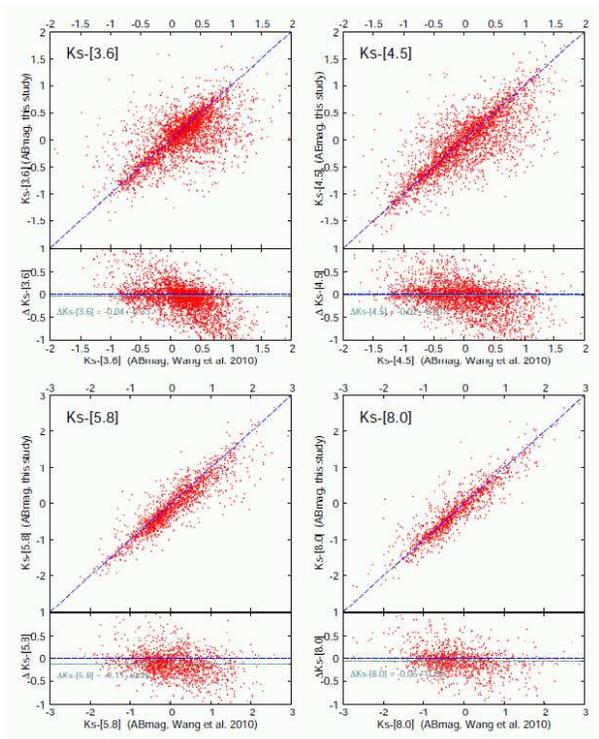}
    %%% \FigureFile(width,height){filename}
  \end{center}
\vspace{-3mm}
  \caption{Comparisons of $K_{s}-[3.6]$, $K_{s}-[4.5]$, $K_{s}-[5.8]$, and 
$K_{s}-[8.0]$ colors between \citet{wan10} and this study. 
}
\label{fig:color3}
\end{figure}

Figure \ref{fig:color3} compares $K_{s}-[3.6]$, $K_{s}-[4.5]$, $K_{s}-[5.8]$, and 
$K_{s}-[8.0]$ between \citet{wan10} and this study.
\citet{wan10} used a deconvolution procedure with a model image made from 
the convolved $K_{s}$-band image, which is similar with that used in 
the subtraction of the contribution from neighboring sources in our 
photometry (Section \ref{sec:color}). On the other hand,  we simply performed 
the aperture photometry after subtracting the contribution from neighboring sources 
and then applied the aperture correction with the surface brightness profiles of 
the $K_{s}$-band images. These colors agree relatively well within the uncertainty, 
although the colors in the MODS catalog tend to be bluer than those by \citet{wan09} 
at relatively red colors, and redder at the bluest colors, especially in 
the $K_{s}-[3.6]$ and $K_{s}-[4.5]$ colors. These systematic differences may be 
due to the differences in the method of the color measurement. 

%%%
% See the manual for the detail.
%%%

\end{document}